\documentclass[]{aa}

\usepackage[varg]{txfonts}
\usepackage{graphicx}
\usepackage{natbib}
\usepackage[dvipsnames]{xcolor}
\usepackage{subfig}
\usepackage{multirow}
\usepackage{hyperref}
\hypersetup{
    colorlinks=true,
    linkcolor=NavyBlue,
    filecolor=NavyBlue,   urlcolor=blue,
    citecolor=NavyBlue
    }

\usepackage[labelfont=bf]{caption}

\begin{document}

\title{Stragglers of the thick disc}

\author{V. Cerqui\inst{}, M. Haywood\inst{}, P. Di Matteo\inst{},
D. Katz\inst{},
F. Royer\inst{}
}

\offprints{V. Cerqui, \email{valeria.cerqui@obspm.fr}}

\institute{GEPI, Observatoire de Paris, PSL Research University, CNRS, Place Jules Janssen, 92195 Meudon, France
}

\authorrunning{V. Cerqui et al.}

\date{}
\abstract{
{Young alpha\text-rich (YAR) stars have been detected in the past as outliers to the local age $\rm-$ [$\alpha$/Fe] relation. These objects are enhanced in $\alpha$\text-elements, but they are apparently younger than typical thick disc stars.}
{}
{Here, we study the global kinematics and chemical properties of YAR giant stars in the APOGEE DR17 survey and show that they have properties similar to those of the standard thick disc stellar population.}
{This leads us to conclude that YAR are rejuvenated thick disc objects, and the most likely explanation is that they are evolved blue stragglers. This is confirmed by their position in the Hertzsprung-Russel diagram (HRD). Extending our selection to dwarfs allowed us to obtain the first general straggler distribution in an HRD of field stars. We also compared the elemental abundances of our sample with those of standard thick disc stars and found that our YAR stars are shifted in oxygen, magnesium, sodium, and the slow neutron-capture element cerium.  
Although we detected no sign of binarity for most objects, the enhancement in cerium may be a signature of a mass transfer from an asymptotic giant branch companion. 
The most massive YAR stars suggest that mass transfer from an evolved star may not be the only plausible formation pathway and that other scenarios, such as collision or coalescence, should be considered.
}
}


\maketitle
\section{Introduction} 
\label{sec:intro}
The majority of disc stars present a trend positing that older stars have higher values of $\rm [\alpha/Fe]$. This can be explained in the context of Galactic chemical evolution: stars born at early times were formed from a gas enriched in $\rm \alpha$\text-elements produced by Type II supernovae (SNe), which dominated the chemical enrichment in the initial epochs. Over time, the production of Fe by Type Ia SNe lowered the level of $\rm [\alpha/Fe]$ \citep{tinsley1979, Matteucci_Greggio_1986}.
The presence of two different stellar populations in the Galactic disc has been confirmed through the presence of two well defined and well separated sequence in the $\rm [\alpha/Fe]-[Fe/H]$ plane \citep[e.g.][]{fuhrmann2011_local, Haywood2013, Recio-Blanco2014, hayden2015}, where the so-called high-$\rm \alpha$ stars are older than the low-$\rm \alpha$ ones \citep[e.g.][]{Haywood2013, bensby2014}. 
Young $\rm \alpha$\text-rich (YAR) stars have been known since at least the study of \citet{Fuhrmann_Bernkopf1999}, who noted the abnormally young age of HR 4657 given its $\rm [\alpha/Fe]$ abundance ratio \citep[see also][]{Fuhrmann2011, Fuhrmann2012}. They stand out as outliers to the general trends. 
For instance, some YAR objects were noted in \citet{Haywood2013}, and discussed further in \citet{Haywood2015}, as outliers of the observed $\rm[\alpha/Fe] - age$ correlation of solar neighbourhood stars. 
Additional works have made use of new asteroseismic measurements and spectroscopic observations to identify YAR objects being part of a population of stars that cannot be explained by standard chemical evolution models. They are found to be $\rm \alpha$\text-enhanced (typically $\rm [\alpha/Fe] > 0.1$ dex) but younger ($\rm age < 6.0$ Gyr) than the typical high-$\rm \alpha$ old thick disc stars, with a typical mass of 1.5 M$_{\odot}$ \citep[e.g.][]{Chiappini2015, martig2015, Jofre2016, Matsuno2018, silva-aguirre2018, Sun2020, Zhang2021, Jofre2022}.
The fraction of YAR stars is estimated to be $\rm \sim 6 \%$ of the $\rm \alpha$\text-rich population \citep{martig2015}.

Until now, two explanations for the origin of YAR stars have been proposed. 
One possible idea relies on star formation events in the region of the bar corotation area \citep{Chiappini2015}. In this view, the gas in this region would be kept isolated for long time, so that YAR stars that formed from this cloud may show the same chemical enrichment of thick disc stars, but are younger in age. In support of this thesis, \citet{Chiappini2015} found the YAR stars of their sample to be located in the inner region of the Galaxy and having dissimilar kinematics from other $\rm \alpha$\text-rich stars.
The second explanation follows the evolved Blue Stragglers scenario. Blue Straggler Stars (BSSs) are believed to be the result of either a stellar merger \citep{Hills-Day1976, Momany2007} or mass transfer in binary systems \citep{McCrea1964, Paczynski1971, Webbink1985}. In both cases, they have experienced an episode of mass acquisition and for this reason, they can be identified as stars bluer and brighter than turn-off stars in clusters.
The measured masses of these stars do not reflect their ages, but the age expected from their high mass is lower than their true age.
Considering this point of view, YAR stars are then thought to be stragglers stars of the thick disc, namely, rejuvenated thick disc stars \citep[e.g.][]{martig2015, Yong2016, Izzard2018}.
A plausible approach to probing this scenario is to search for the evidence of mass transfer due to binary evolution in the spotted YAR sample of stars.
\citet{Jofre2016} and the follow-up work \citep{Jofre2022}, for instance, made a radial velocity monitoring campaign using the HERMES spectrograph to evaluate whether YAR stars were or were not part of binary systems. 
In particular, in \citet{Jofre2022}, the authors concluded that YAR stars are very likely to be products of mass transfer, thus they would effectively not be young.
Moreover, if YAR objects are rejuvenated thick disc stars, their kinematics and spatial distribution in the Galaxy should be the same as those of this population. 
This has been verified by recent works, as in \citet{Sun2020} and \citet{Zhang2021}, where the LAMOST dataset was used to compare the characteristics of the thick  and thin disc samples with the identified YAR population.

In this paper, we aim to provide stronger constraints on the thick disc straggler scenario by examining and comparing the global properties of the YAR objects found in the APOGEE data release 17 (DR17) \citep{Abdurro'uf_apogee}.
To achieve this goal, we utilised the age estimates from the value added catalogue $\rm apogee\_astroNN-DR17$ (hereafter, the astroNN catalogue; \cite{Leung_Bovy2019})\footnote{The astroNN python package is  available at \url{https://github.com/henrysky/astroNN}.}. These age estimates are derived from the $\rm [C/N]$ ratio-mass relationship, making them a useful tool for examining objects that are believed to have formed due to an increase in the parent star's mass, such as YAR stars. It is worth noting that the astroNN catalogue does not provide direct stellar mass estimates, but they are reflected in the age estimates.
In the following section, we describe how we selected our YAR candidates. Section \ref{sec:Gen_prop} presents the general properties of our sample, including the Hertzsprung-Russel diagram (HRD), the metallicity and $\rm \alpha$ abundances, and the kinematic properties of our YAR stars. In Section \ref{sec:accreted}, we look for possible YAR accreted candidates, while in Section \ref{sec:abundances} we discuss the individual chemical patterns of our stars. In Section \ref{sec:general_YAR}, we extend our selection to all gravities.  We discuss our results in Section \ref{sec:discussion} and present our conclusions in Section \ref{conclusions}.

\section{Data}
\label{sec:data}
We used the APOGEE atmospheric parameters and stellar abundances from data release 17 ($\rm allStarLite-dr17-synspec\text{\textunderscore} rev1$). 
\begin{figure}
\includegraphics[width=\hsize]{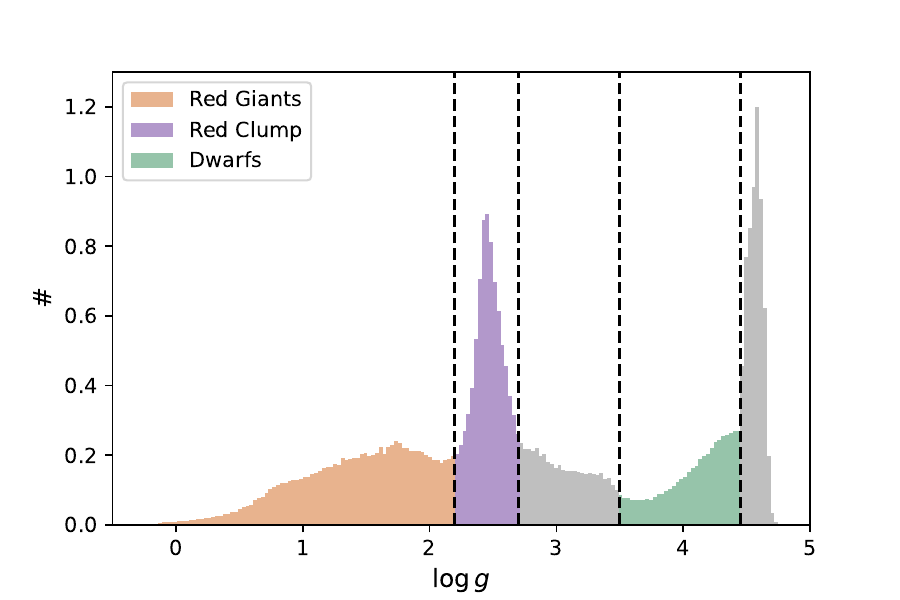}
\includegraphics[width=\hsize]{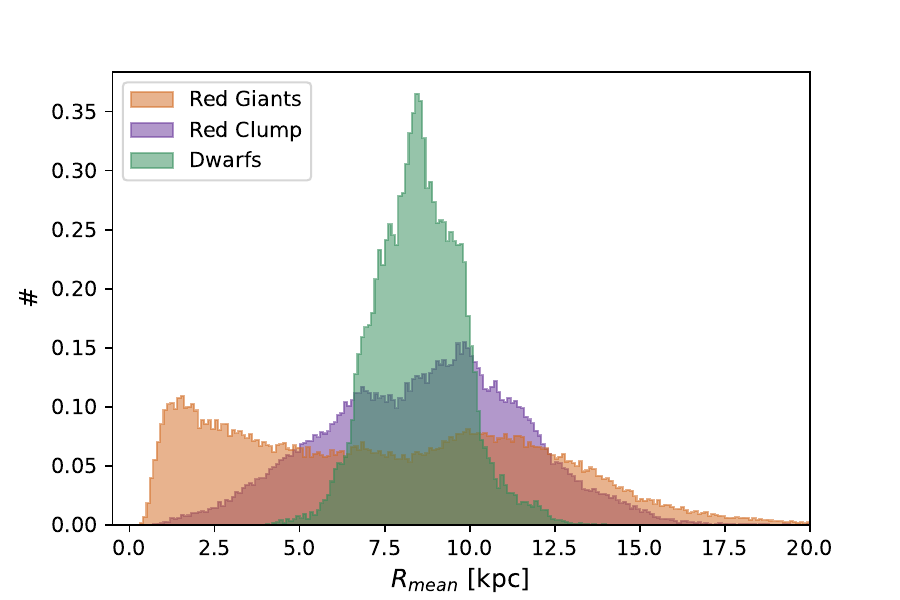}
\caption{Distributions of surface gravity and mean radius of APOGEE DR17 flag selected stars. The $\rm \log g$ normalised histogram is shown at the top. The dashed vertical lines represent the thresholds chosen to select stars in three $\rm \log g$ intervals (red giants: $\rm \log g < 2.2$, in orange; red clump stars: $\rm 2.2 < \log g < 2.7$, in purple; dwarfs: $\rm 3.5 < \log g < 4.45$, in green). The mean orbital radii density histogram of the selected stars separated into the three $\rm \log g$ intervals is shown at the bottom, with the same legend as in the top panel.}
\label{type_stars}
\end{figure}
\begin{figure}
{\includegraphics[width=\hsize]{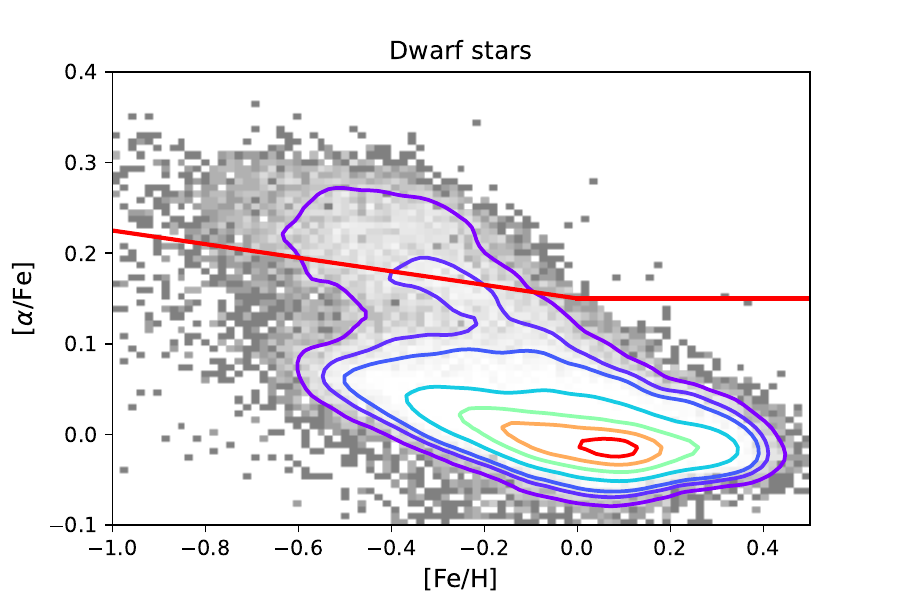}
\includegraphics[width=\hsize]{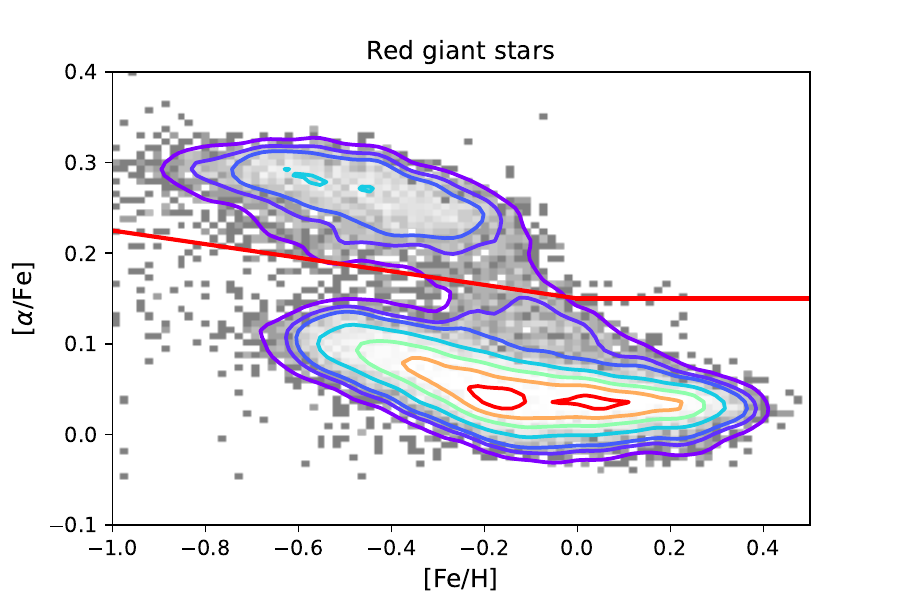}
\includegraphics[width=\hsize]{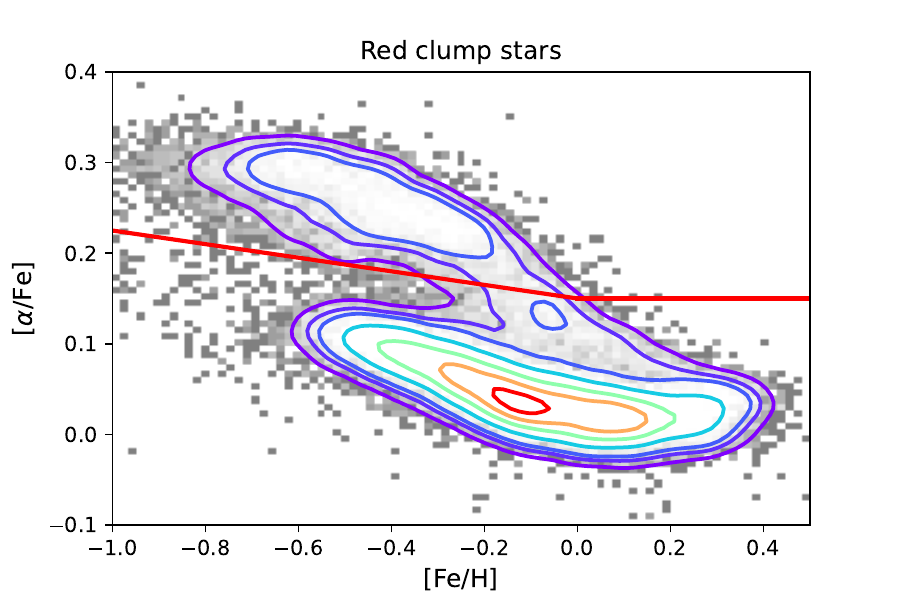}}
\caption{$\rm [Fe/H]-[\alpha/Fe]$ distributions for dwarfs, red giants and clump stars within 3 kpc from the Sun. The red lines in each panel illustrate our definition of low and high-$\rm \alpha$ stars. The coloured lines represent stellar
density iso-contours: $\rm 90 \%$, $\rm 70\%$, $\rm 50\%$, $\rm 30\%$, $\rm 15\%$, $\rm 10\%,$ and $\rm 5\%$ of the peak density.
}
\label{alphafeh}
\end{figure}
When using abundances, we selected 364,605 stars flagged as $\rm ASPCAPFLAG\, bit\, 23 == 0$ (no flag for BAD overall for stars), $\rm EXTRATARG == 0$ (main survey stars) and $\rm SNR > 50$. 
The resulting selection contains stars with widely different atmospheric parameters, as illustrated by the distribution of $\rm \log g$ for the whole sample (see Fig. \ref{type_stars}). 
We divided the sample into three groups according to $\rm \log g$: red giants at $\rm \log g < 2.2$, clump giants at $\rm 2.2 < \log g < 2.7$, and dwarfs at $\rm 3.5 < \log g < 4.45$ (top panel of Fig.~\ref{type_stars}).
As illustrated in the bottom panel of Fig.~\ref{type_stars}, the stars in these three groups sample different ranges of mean radius $\rm R_{mean}$, defined as the mean of the galactic pericenter distance and the galactic apocenter distance. They also have chemical patterns that differ significantly (Fig.~\ref{alphafeh}) for reasons that are possibly related to intrinsic age and mass distributions of each of these type of stars, but also to the spectroscopic analysis on different type of stars. 

\subsection{Age estimates}\label{subsec:age}

We used age and orbital parameter estimates provided by the astroNN catalogue. 
Stellar ages in astroNN were estimated, as explained in \citet{Mackereth2019}, using a Bayesian neural network trained on asteroseismic ages. 
The age estimates are based on the $\rm [C/N]$ abundance ratio. This ratio is modified during the first dredge-up, when the star starts the red giant branch (RGB) phase. The intensity of the mixing (hence, the amplitude of the change of the $\rm [C/N$] ratio) is linked to the depth of the first dredge-up, which is itself linked to the mass of the star. For this correlation to appear, it is necessary for the star to have passed the first dredge-up, after leaving the subgiant phase; therefore, this is not expected to apply to the case of dwarfs. The $\rm [C/N]$ abundance ratio has been utilised in numerous works in literature for dating giant stars, with the first being the work by \cite{Masseron_Gilmore2015}.
The median uncertainty on ages reported in \citet{Mackereth2019} from applying their procedure to the DR14 is $\rm 30\%$, while these authors also caution that ages above 10 Gyr are probably underestimated, by as much as 3.5 Gyr.

\begin{figure}
\includegraphics[width=\hsize]{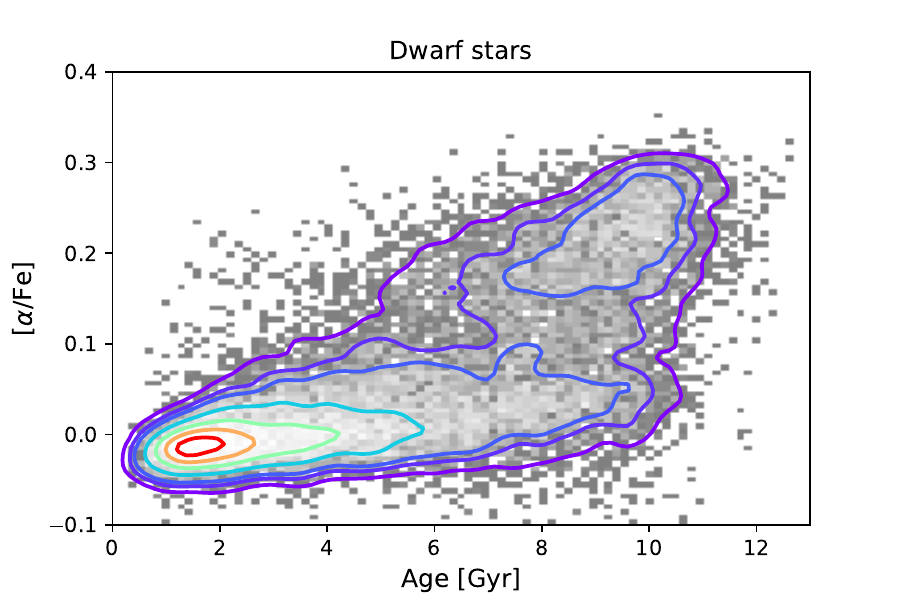}
\includegraphics[width=\hsize]{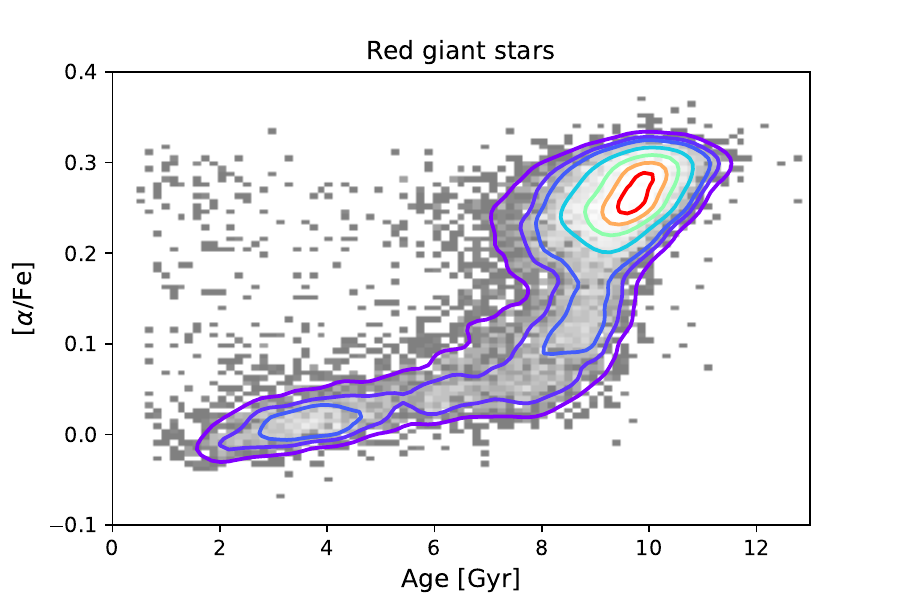}
\includegraphics[width=\hsize]{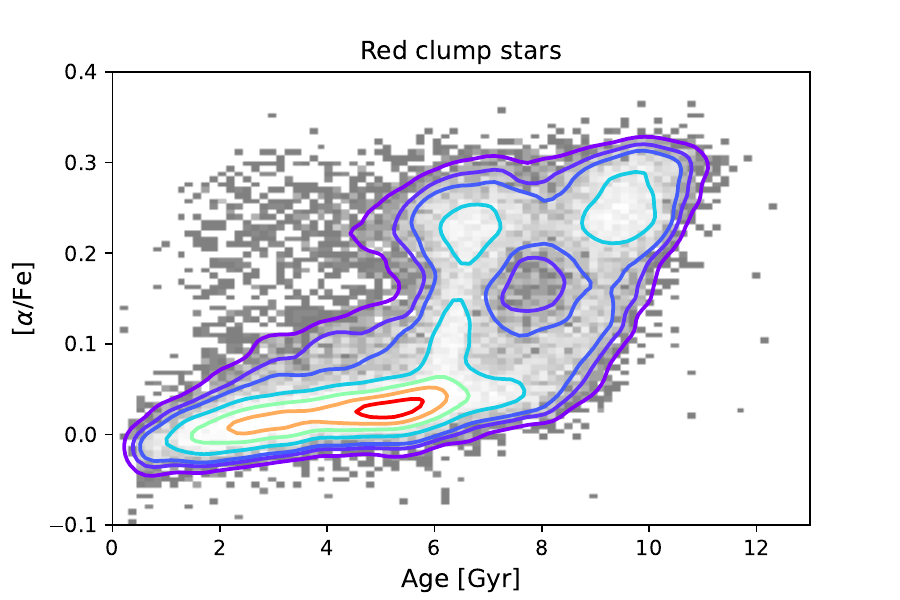}
\caption{$\rm Age - [\alpha/Fe]$ distributions for dwarfs, giants, and clump stars, as defined in Section \ref{sec:data} at $\rm R_{mean} < 8$ kpc from the Galactic centre. Red giants and red clump stars are selected to have an error on age below 2 Gyr. The coloured lines represent stellar
density iso-contours: $\rm 90 \%$, $\rm 70\%$, $\rm 50\%$, $\rm 30\%$, $\rm 15\%$, $\rm 10\%,$ and $\rm 5\%$ of the peak density.}
\label{age_alpha_type_stars}
\end{figure}

Figure \ref{age_alpha_type_stars} shows the $\rm age-[\alpha/Fe]$ distributions for the three types of stars selected to be within 8 kpc from the Galactic centre. 
This figure illustrates that the age scale for the different types of stars is different, with the break between thin and thick disc stars occurring at 6, 8, and around 9-10 Gyr,  respectively, for red clump stars, giants, and dwarfs; this suggests $\rm \sim$ 2 Gyr offsets between the different age scales.
The top plot shows that $\rm \alpha$\text-rich dwarf stars were correctly identified as old objects because, in fact, the neural network model probably learned that the $\rm \alpha$\text-rich stars are old stars in general.
The $\rm age - [\alpha/Fe]$ relation for clump stars (bottom plot) shows two high\text-$\rm \alpha$ sequences. This feature is caused by the possible contamination by red giant stars of the red clump-selected sample.
Because of the difference in the age scales for red giants and red clump stars, and the fact that the former cover a wider range in mean radii, we focus this study on red giants. We comment on clump and dwarf YAR stars in Section \ref{sec:general_YAR}.

For binaries that are stragglers, it is expected that the change in the $\rm [C/N]$ ratio will depend on when the mass transfer or merger occurs. If the straggler forms before the start of the ascent of the red giant branch, then the $\rm [C/N]$ ratio is expected to decrease as in a normal massive star. If it forms later, $\rm [C/N]  $ may not be affected by the formation of the straggler and it will not be correlated with the mass of the straggler. This implies that thick disc stars detected as apparently young (and truly massive) on the basis of their $\rm [C/N]$ ratio will represent only a fraction of stars of the thick disc that have acquired mass due to the straggler mechanism. An unknown fraction of stars will remain undetected as stragglers because their $\rm [C/N]$ ratio will remain unaffected by the mass transfer or merger of the system.
In Section \ref{subsec:dhr}, we show that the objects selected as YAR on the basis of the age estimates drawn from the $\rm [C/N]$ ratio are effectively more massive than standard thick disc stars.

\subsection{The young $\rm \alpha$\text-rich (YAR) sample}
\label{subsec: YAR-defin}
\begin{figure*}
\label{fig: age-alpha}
\centering
\includegraphics[width=\hsize]{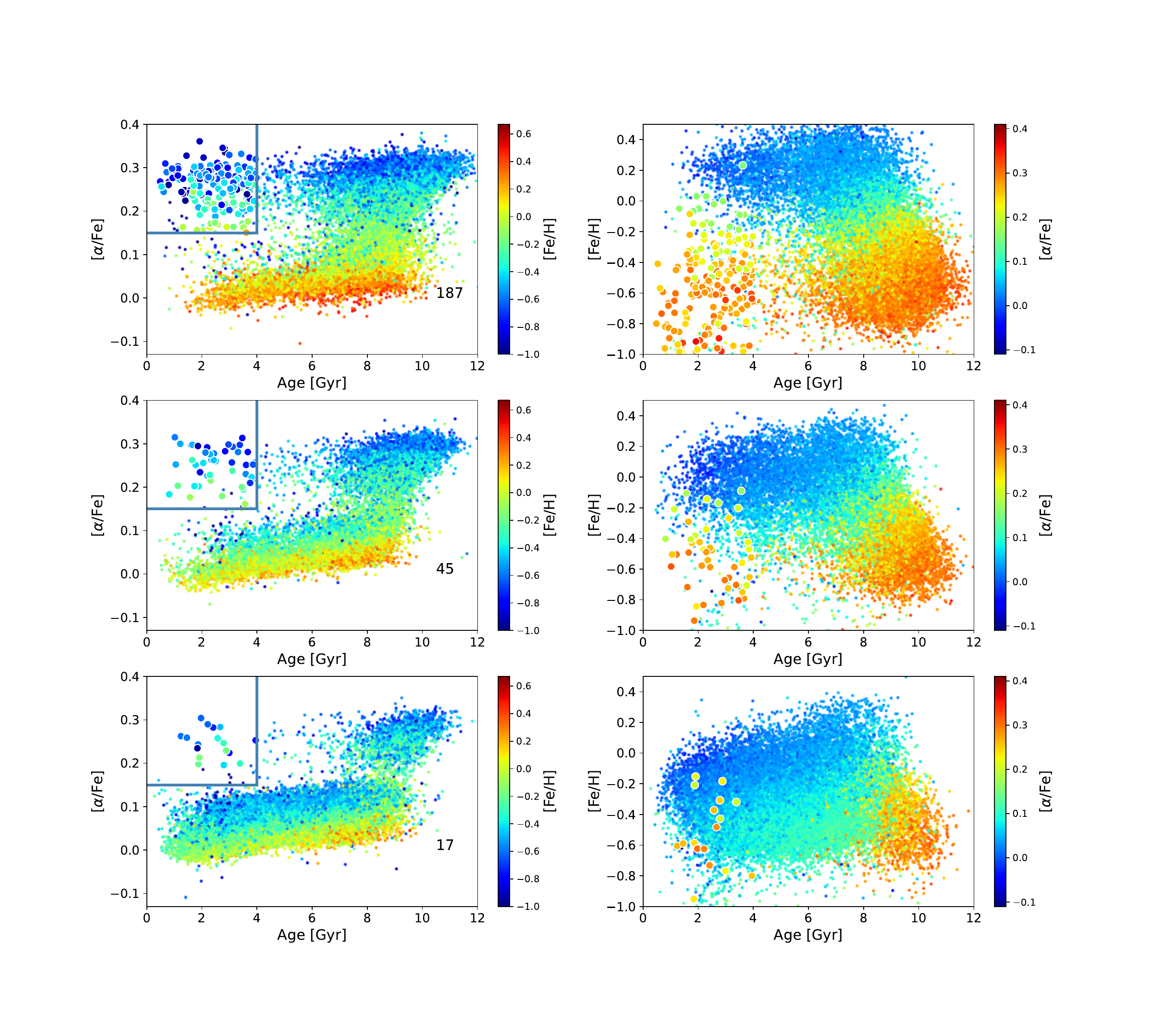}
\caption{Sample used in this study divided in three separate intervals of mean radius:  0-6 kpc (\textit{top panels}), 6-9 kpc (\textit{middle panels}), and 9-20 kpc (\textit{bottom panels}). The number of stars above the line at $\rm [\alpha/Fe]=0.15$ and younger than 4 Gyr is indicated on each plot. All the stars in this figure were selected to have an error on age below 3 Gyr.}
\label{fig:selections}
\end{figure*}

We defined our sample of YAR by selecting bright red giants ($\rm \log g < 2.2$) with age less than 4 Gyr; age error less than 3 Gyr; $[\rm \alpha/Fe] > 0.15$ dex at $\rm [Fe/H] > 0$ dex and $\rm [\alpha/Fe] > -0.075\times[Fe/H] + 0.15$ dex below solar metallicity, so that they are located above the red line in the central upper panel of Fig.~\ref{alphafeh} (which divides the high\text-$\rm \alpha$ thick disc sequence from the  low\text-$\rm \alpha$ thin disc sequence).
 In this way, we selected  our sample of 249 stars.
Figure \ref{fig:selections} shows our sample divided into three different intervals of mean radii from the Galactic centre: 0\text-6, 6\text-9, and 9\text-20~kpc.
The number of YAR stars increases sharply from $\rm R_{mean} > 9$ to  $\rm R_{mean} < 6$~kpc, as expected if these objects are dominated by thick disc stars. 
The precise fractional number of stars of both the thick disc and YAR samples for different rangesof the radius is listed in Table \ref{table: frac_n}.
\begin{table}
\caption{Fractional number of stars for thick disc and YAR samples in different mean radii ranges.}             
\label{table: frac_n}      
\centering                          
\begin{tabular}{c c c c}        
\hline\hline                 
 & $\rm N_{0-6}/N_{6-9}$ &  $\rm N_{6-9}/N_{9-20}$ \\ 
\hline                        
 Thick disc   &   $\rm 2.98^{max=3.04}_{min=2.92}$& $\rm 2.66^{max=2.75}_{min=2.56}$\\
 YAR&  $\rm 4.16^{max=5.24}_{min=3.35}$& $\rm 2.65^{max=4.02}_{min=1.81}$  \\
 \hline                                  
\end{tabular}
\tablefoot{The second column represents the fraction of star within 6 kpc from the galactic centre over the stars in the range 6-9 kpc. The third column show the analogous quantity but for 6-9 kpc and 9-20 kpc ranges. The fraction can be minimised or maximised (min and max) taking into account the Poisson uncertainties of $\rm N_{0-6}$, $\rm N_{6-9}$, and $\rm N_{9-20}$.}
\end{table}
The choice of the limit on the age error impacts the number of low-metallicity stars. Selecting stars with errors on their ages lower than 3 Gyr essentially removes  all stars below metallicity of -0.9 dex, except for the youngest objects, with ages lower than 4 Gyr (visible in the age-metallicity distribution plots). 
Setting limits at ages lower than 4 Gyr and age errors lower than 3 Gyr comes from a compromise between stars that can  reasonably be considered as bona\text\  fide young alpha\text-rich objects and the total number of objects.  
The mean error on age of the YAR is 2.08 Gyr.

We made use of the Gaia DR3 data \citep{Gaia_DR3}  to investigate the binarity of our YAR sample.
In particular, we explored the renormalised unit weight error (RUWE)  included in the \texttt{gaia\textunderscore source} table. Indeed, this parameter expresses an indication of the astrometric solution quality of a source and can then give a good hint of a star being in an unresolved multiple system (this affecting the goodness and the uncertainty of the astrometric solution). 
Only 9 stars out of the 249 (about 4$\rm \%$) in our sample display a value of  $\rm \texttt{ruwe} > 1.4$. This is the typical value above which the parallax measurement of a source can be considered less reliable \citep{Lindegren2021}. For this reason, we can expect sources characterised by $\rm \texttt{ruwe} > 1.4$ as likely multiple systems.
In addition, just two YAR stars in the sample are catalogued as binaries (specifically, a single-lined spectroscopic binary and a combined astrometric + single lined spectroscopic orbital model binary) by the
\texttt{nss\textunderscore two\textunderscore body\textunderscore orbit} catalogue of the non-single stars (NSS) tables of Gaia DR3 \citep{Gaia_binary}. 
A possible reason for such a low percentage of multiple systems among our YAR sample can be the possible small orbital separation of the system detected, which should not significantly impact the astrometric solutions \citep[as suggested in][]{Kounkel2021}. In addition, as we discuss in Section \ref{sec:discussion}, straggler stars can be the product of a merging or a coalescence phenomenon. We note that the YAR object formed in this case is a single star.

The \texttt{source} table of Gaia DR3 provides as well two variability indicators based on radial velocity time series: the radial velocity renormalised goodness of fit (\texttt{rv\textunderscore renormalised\textunderscore gof}) and the P-value for constancy based on a chi-squared criterion (\texttt{rv\textunderscore chisq\textunderscore pvalue}). They are both limited to stars brighter than $\rm G_{RVS}=12$.
The selection that we adopted (suggested in the Gaia DR3 documentation) to find potential radial velocity variable stars also invokes  the total number of epochs used to obtain the radial velocity (\texttt{rv\textunderscore nb\textunderscore transits}), namely: \texttt{rv\textunderscore renormalised\textunderscore gof} $>$ 4\, \text{AND}\, \texttt{rv\textunderscore chisq\textunderscore pvalue} $<$ 0.01 \text{AND}\, \texttt{rv\textunderscore nb\textunderscore transits} $\geq$ 10.
12 YAR stars out of 249 (about 5$\rm \%$) are classified as variable with this method, with 4 of them having $\rm \texttt{ruwe} > 1.4$.

For the  most complete binarity analysis possible, we finally explored the radial velocity scatter (corresponding to the parameter \texttt{vscatter}) from APOGEE DR17. We catalogued stars as likely binaries when $\rm \texttt{vscatter} > 1\, km/s$, as suggested in the APOGEE DR17 documentation. In fact, the histogram of the radial velocity scatter (for the complete sample of APOGEE DR17) peaks at around 70 $\rm m/s$ and presents a long tail at higher scatter values, probably due to variability from stellar binaries. Thus, 1 $\rm km/s$ represents a good threshold compared to the peak of the distribution.
The 21 YAR stars (approximately 8$\rm \%$) are the outcome of this selection: 3 of them are classified as variable stars through the Gaia DR3 radial velocity and 1 of them has $\rm \texttt{ruwe} > 1.4$.
In total, from these investigations, 35 of the 249 YAR stars show at least one sign of binarity between RUWE, Gaia DR3 NSS solution, Gaia DR3 radial velocity time series and APOGEE DR17 radial velocity scatter and only 1 star shows signs of binarity in all diagnostics.

We selected a sample of thick disc stars as a reference population, defined here as being at $[\rm \alpha/Fe] > 0.15$ at $\rm [Fe/H] > 0$ and $\rm [\alpha/Fe] > - 0.075 \times [Fe/H] + 0.15$ below solar metallicity and at ages greater than 4 Gyr. It contains 24,357 stars.

\section{General properties}
\label{sec:Gen_prop}
We go on to study the HRD, metallicity, $\rm \alpha$\text-abundance, and kinematics of our sample and compare it with the thick disc.

\subsection{HRD}\label{subsec:dhr}

\begin{figure*}
\resizebox{\hsize}{!}
{\includegraphics[]{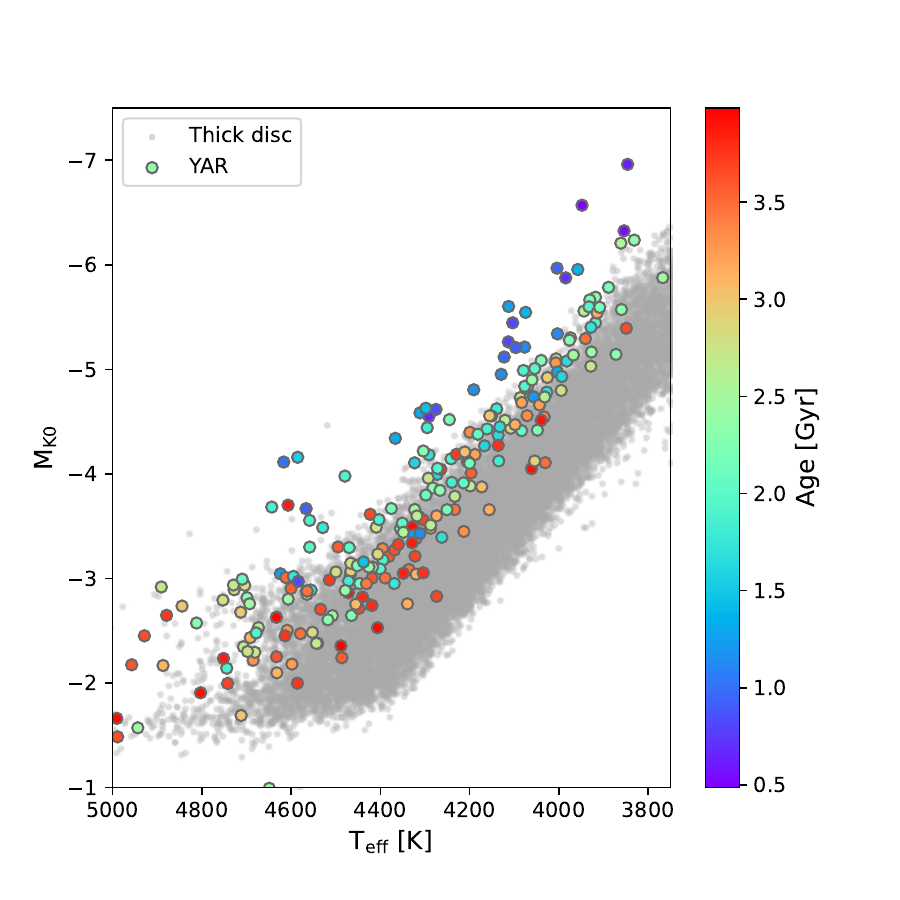}\includegraphics[]{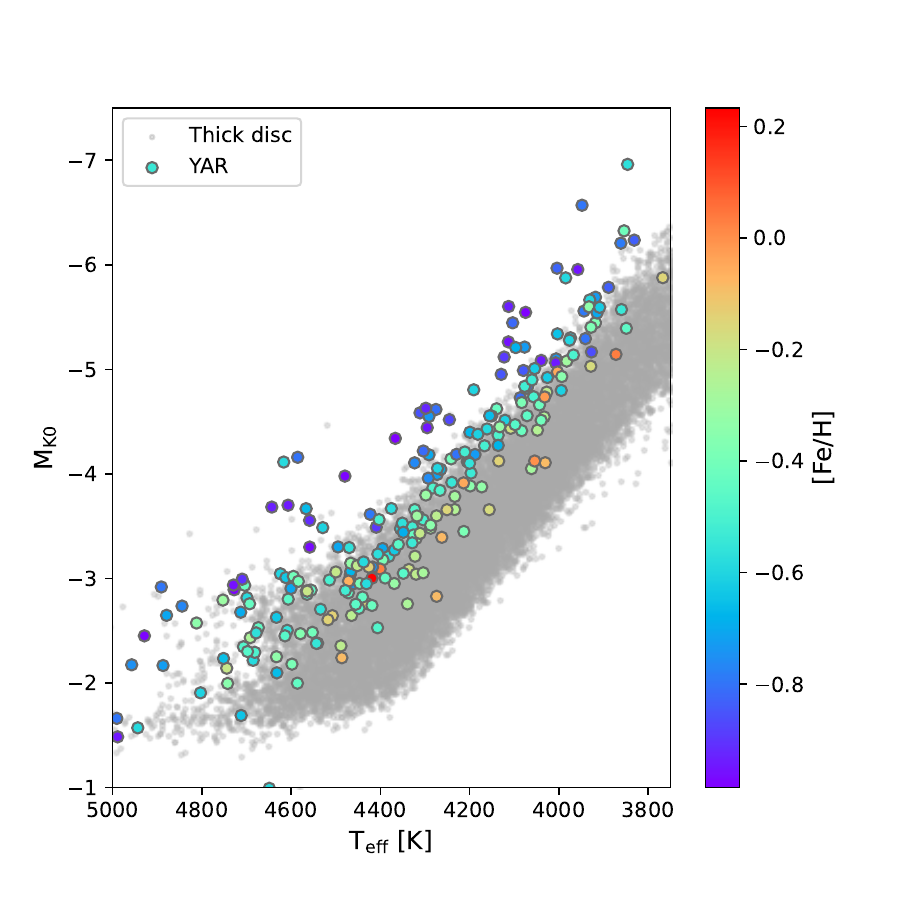}}
\caption{HRD of the YAR stars with $\rm \log g < 2.2$ and $\rm [Fe/H] > -1$. The colour of the points is coding age (\textit{left plot}) and metallicity (\textit{right plot}). In both the panels the small grey dots in background represent the thick disc reference sample (details of the selection are given in Subsection \ref{subsec: YAR-defin}).}
\label{fig: dhr}
\end{figure*}

Figure~\ref{fig: dhr} shows the (T$_{\rm eff}$, M$_{\rm K0}$) HRD of the 249 YAR stars of our sample colour-coded by age (left panel) and metallicity (right panel) of the stars. Background grey points are thick disc red giant stars over the same metallicity interval.  The
M$_{\rm K0}$ magnitudes were obtained using the \texttt{AK$\text{\textunderscore}$TARG} estimate of extinction given in the APOGEE catalogue to correct the K band magnitudes, and the distance was taken from the astroNN catalogue.
YAR stars are located to the left of the background distribution, as expected if they represent the evolution of more massive stars. We note that (as expected from the left panel of Fig.~\ref{fig: dhr}), within the population of YAR stars, the most massive objects (as measured by their young age) are hotter than the less massive ones. The youngest (and most massive) objects and the most metal-poor occupy the upper left part of the sequence, as expected.  Inversely, the oldest (age greater than 3 Gyr) and most metal-rich stars are dominant on the bottom-right part of the sequence.

\begin{figure*}
\resizebox{\hsize}{!}
{\includegraphics[]{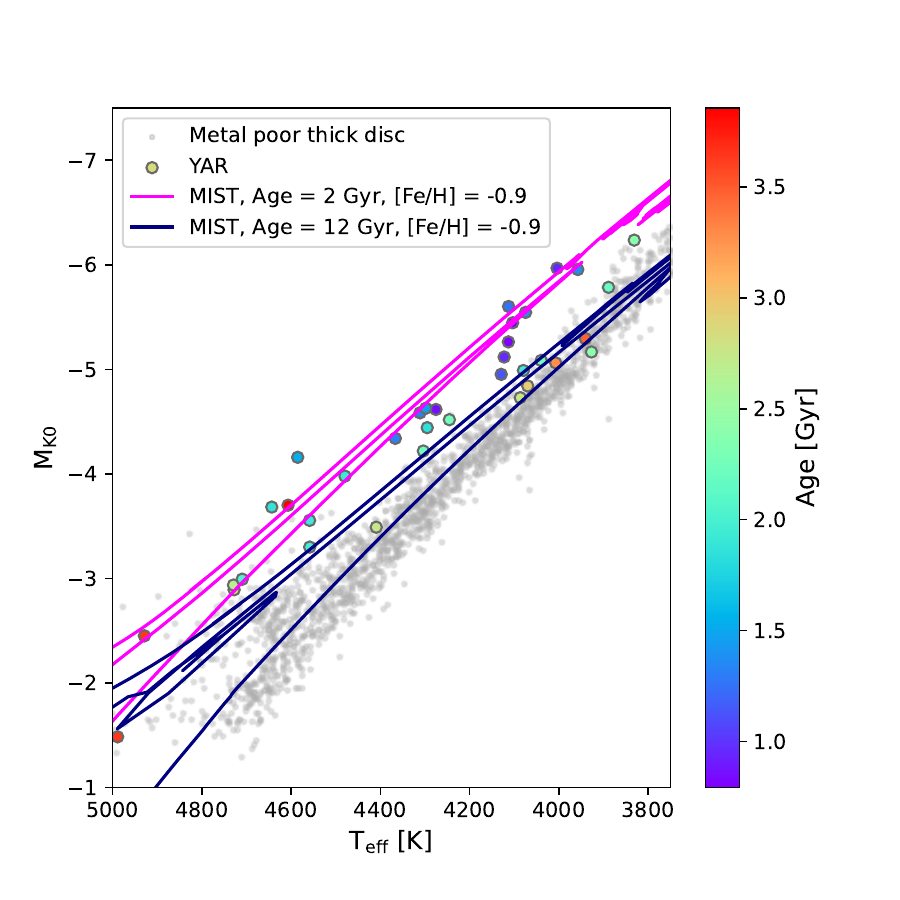}\includegraphics[]{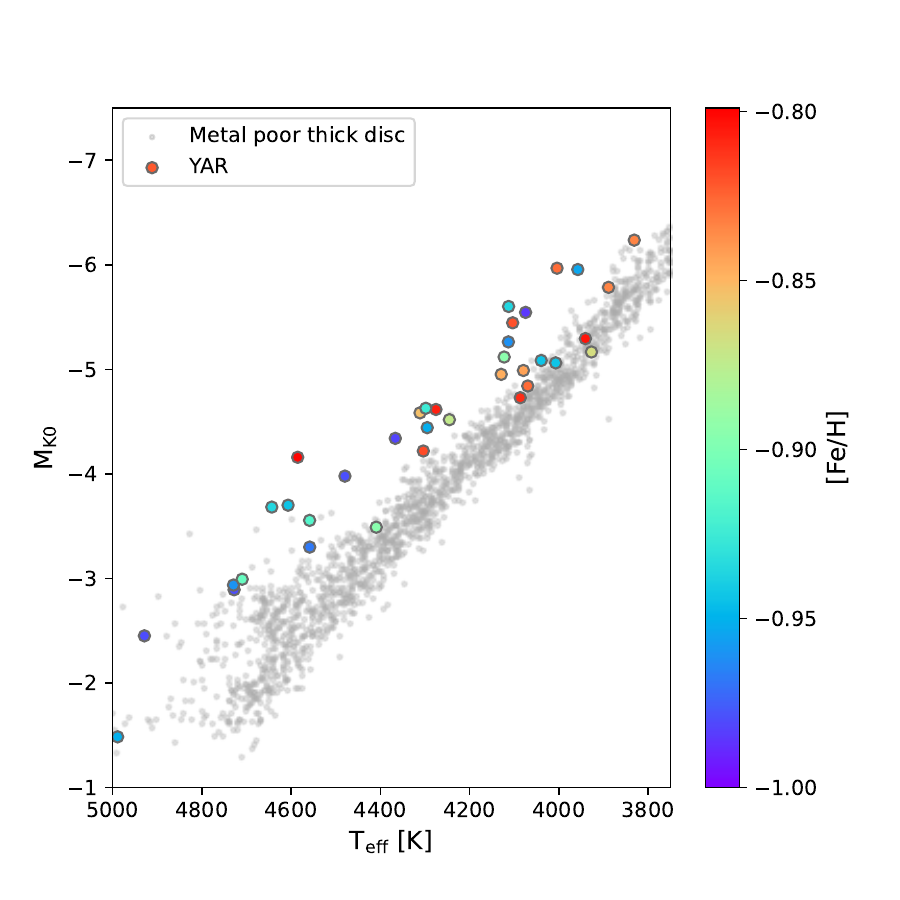}}
\caption{HRD of the YAR stars with $\rm \log g < 2.2$ and $\rm -1 < [Fe/H] < -0.8$ dex. The colour of the points is coding age (\textit{left plot}) and metallicity (\textit{right plot}). MIST isochrones at 2 and 12 Gyr are also plotted on the left plot in fuchsia and dark blue, respectively. In both the panels, the small grey dots in background represent the thick disc reference sample (details of the selection are given in Subsection \ref{subsec: YAR-defin}) restricted to $\rm [Fe/H] < -0.8$.}
\label{fig: dhrpoor}
\end{figure*}

In Fig.~\ref{fig: dhrpoor}, we show the analogous HRD of Fig.~\ref{fig: dhr}, restricting the metallicity interval to $\rm -1 < [Fe/H] < -0.8$ dex, to focus primarily on the effect of age (or mass). The figure illustrates that YAR stars are well detached from the background population at the same metallicity, due to the increase in mass producing apparently younger objects. This is clear when looking at the isochrones in the left panel of the figure. The blue and fuchsia lines represent   old (age = 12 Gyr, approximately the age of the thick disc) and young (age = 2 Gyr) metal poor ($\rm
[Fe/H] = -0.9$ dex) isochrones from  MESA Isochrones and Stellar Tracks (MIST)\footnote{\url{https://waps.cfa.harvard.edu/MIST/}}, respectively. The position of the two sequences reflects their age difference and coincides with the YAR and reference samples.
\begin{figure}
\includegraphics[width=\hsize]{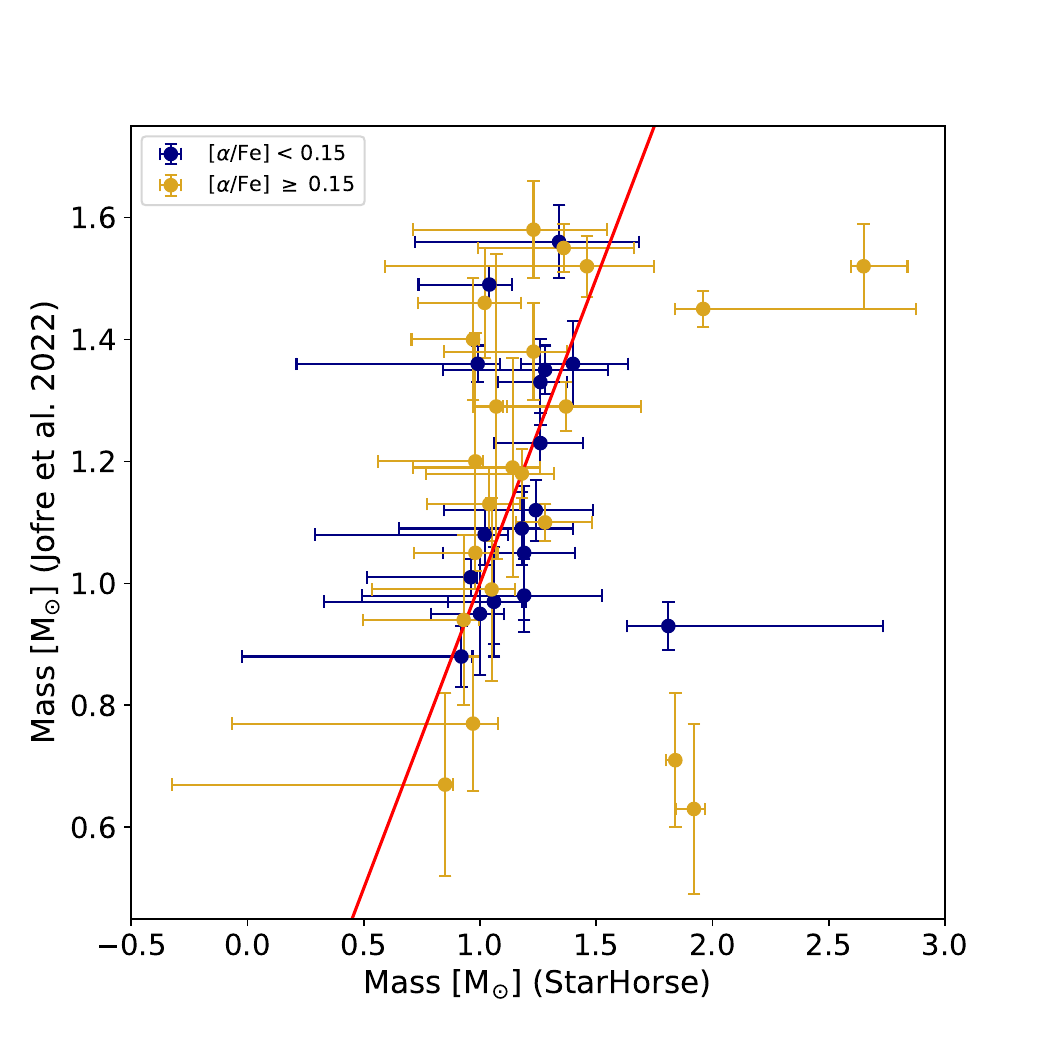}
 \caption{Comparison between asteroseismic mass measurements taken from  table 1 in \citet{Jofre2022} and StarHorse. We distinguish between stars characterised from $\rm [\alpha/Fe] \geq 0.15$ dex (gold dots) and from $\rm [\alpha/Fe] < 0.15$ dex (navy dots). The respective error bars are plotted. The red line is the bisector and represents the $\rm 1:1$ relation.}
\label{fig:jofre22mass}
\end{figure}

To confirm these results, we searched for mass estimates of our stars. These are not provided in the astroNN catalogue, so we took them from the independent StarHorse\footnote{\url{https://data.sdss.org/sas/dr17/env/APOGEE_STARHORSE/}} value-added catalogue. We caution that YAR mass estimates are made as if these stars were normal stars and only reflect their position in the HRD. If these objects are formed through mass transfer or collision, it is possible that their mass is related a bit differently to the effective temperature and luminosity, as compared to normal stars. We do, however, expect this difference to be small \citep{Glebbeek2008, Sills2015} and we view these mass estimates as useful indicator of the true mass of these objects.
In addition, we validated the StarHorse mass determinations by comparing them with  asteroseismic masses. Figure~\ref{fig:jofre22mass} shows the APOKASC-3 asteroseismic mass determinations taken from \citet{Jofre2022} versus the StarHorse estimates. The figure shows that the two datasets are consistent, with only 5 stars being severe outliers of the $\rm 1:1$ relation, highlighted by the red line in the plot. 

\begin{figure}
\centering
\includegraphics[width=\hsize]{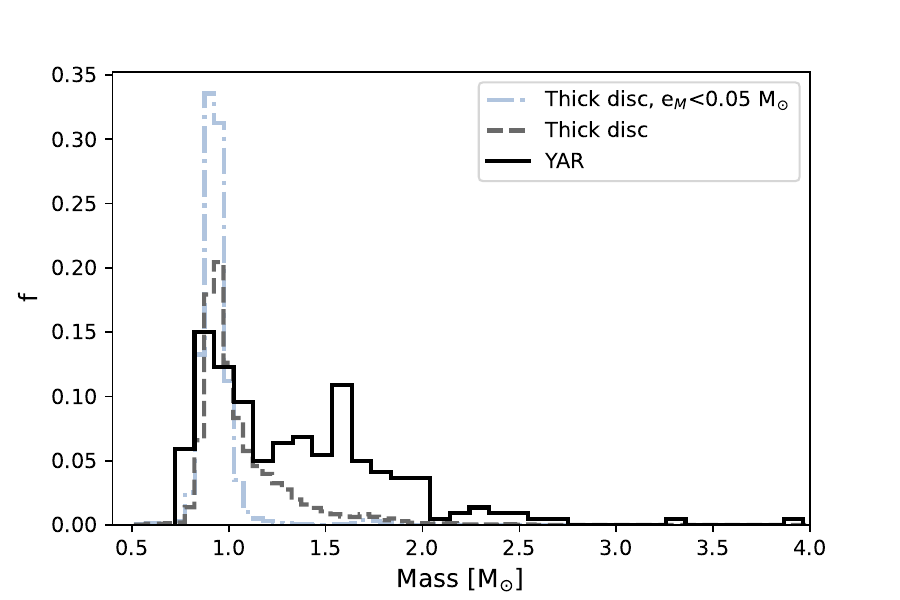}
\caption{Mass distribution function of the YAR (in black solid) and the thick disc (in dark grey dashed, and light grey dash-dotted) samples. $\rm f = N_i/N_{tot}$ represents the fractional number density where $\rm N_i$ is the number of stars in each bin of mass and $\rm N_{tot}$ is the total number of stars of the sample. The light grey dash-dotted histogram restricts the thick disc sample to stars with mass errors less than 0.05 $\rm M_{\odot}$.}
\label{fig:Mass_distrib}
\end{figure}

Figure~\ref{fig:Mass_distrib} shows the mass distribution of the YAR stars and thick disc sample. Mass estimates are available for 220 of our YAR stars, that is, for around 88$\%$ of the total YAR sample.
The thick disc reference sample has a mass distribution peaked at $\rm 0.9-0.95\, M_{\odot}$ while YAR stars sample a wider range of masses, with a significant number of stars with masses up to about 1 $\rm M_{\odot}$ higher than the thick disc stars. This result is in agreement with \citet{Zhang2021}, and, together with the location of YAR stars in the HRD,  confirms that many of the stars in our sample of YAR stars are more massive than standard thick disc objects.

\subsection{Metallicity, C\text-, N\text-, and $\alpha$\text-abundance}
\label{sub: metallicity-alpha}
We now study the metallicity, C\text-, N\text-, and $\rm \alpha$\text-abundance  of our YAR objects and compare them with our thick disc reference sample, defined in  Section \ref{subsec: YAR-defin}.
\begin{figure*}
\resizebox{\hsize}{!}
{\includegraphics[]{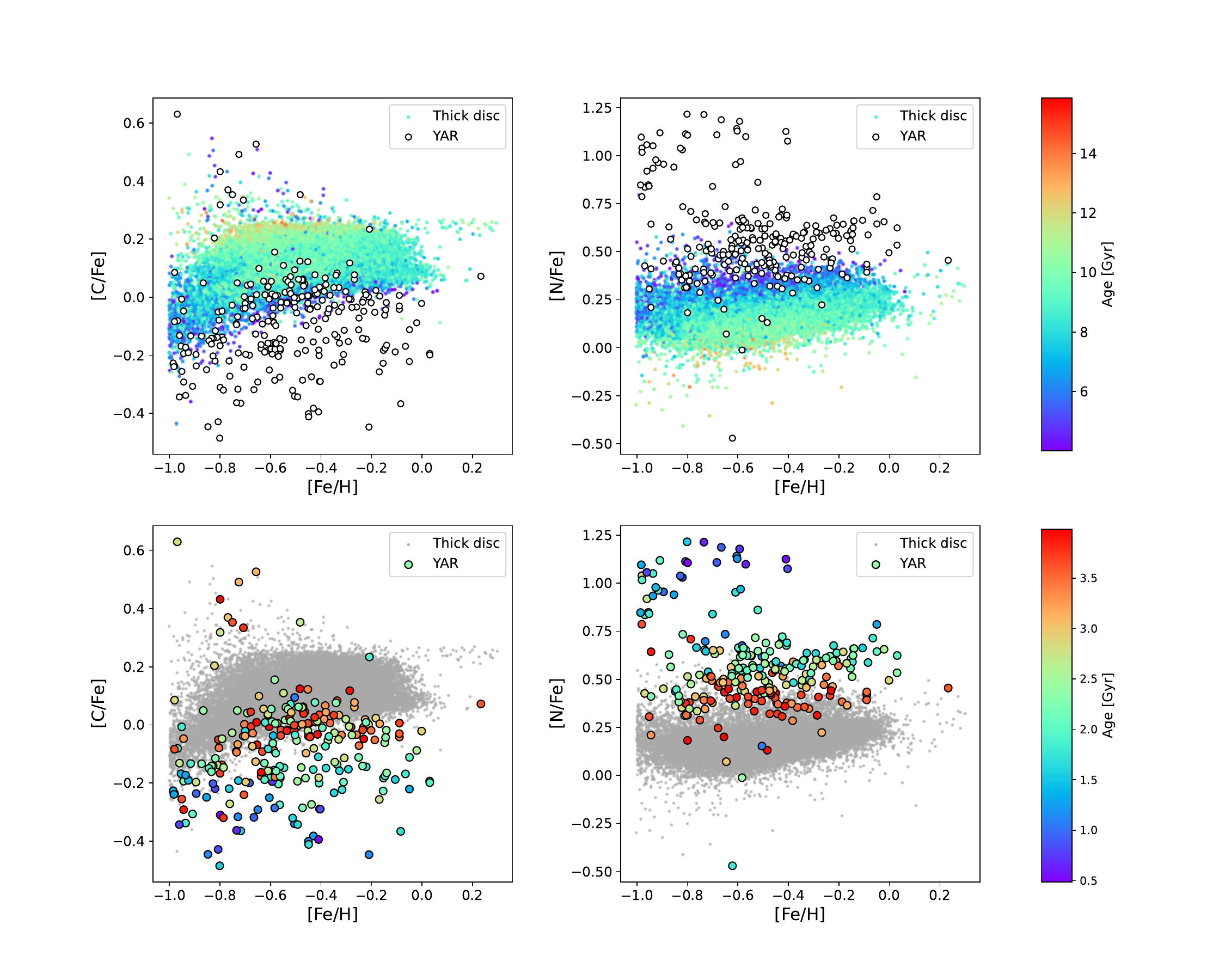}}
\caption{Carbon and nitrogen distributions as a function of metallicity for thick disc and YAR stars. 
\textit{Top plots}: thick disc stars are colour-coded according to their age while YAR stars are shown as black empty circles. \textit{Bottom plots}: YAR stars are colour-coded according to their ages,  while the thick disc sample is shown as grey points.}
\label{fig: CN_thick_YAR}
\end{figure*}

\begin{figure*}
\resizebox{\hsize}{!}
{\includegraphics[width=10cm]{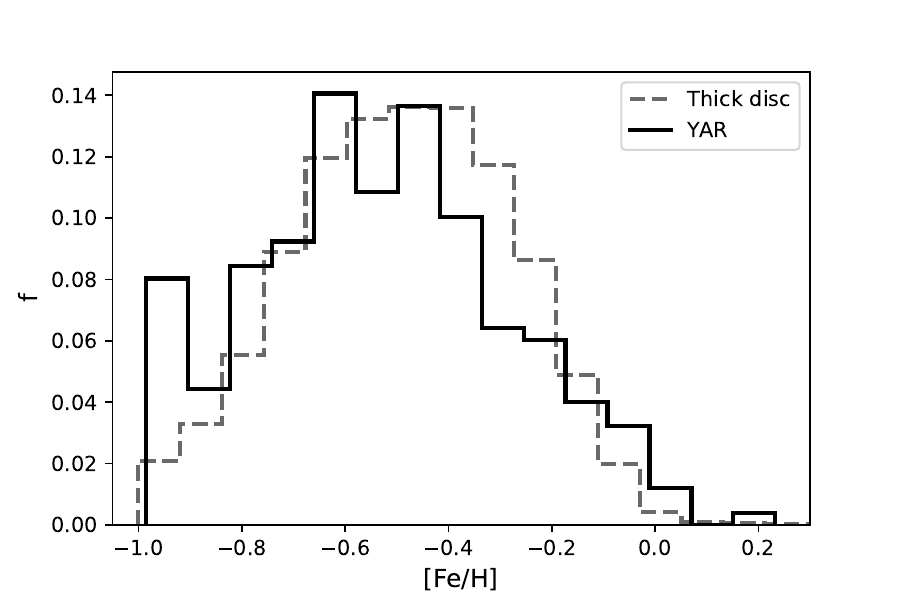}
\includegraphics[width=10cm]{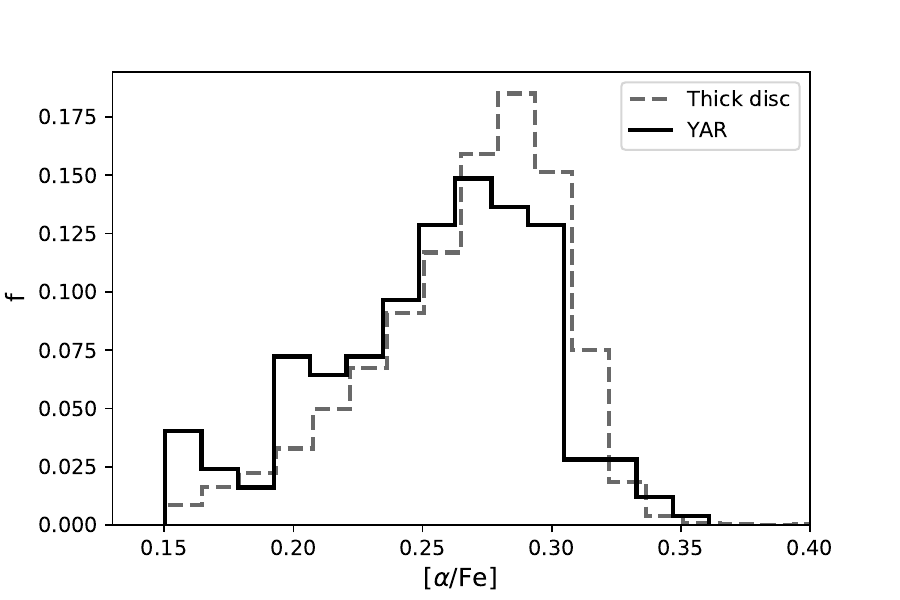}}
\resizebox{\hsize}{!}
{\includegraphics[width=10cm]{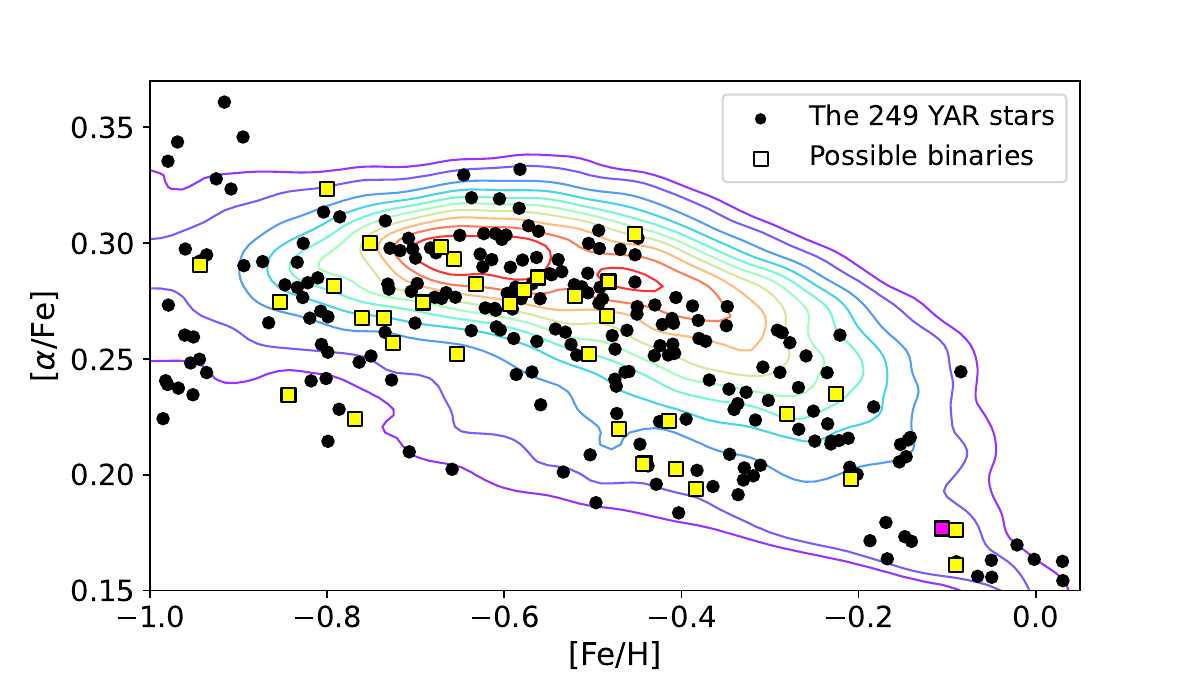}
\includegraphics[width=10cm]{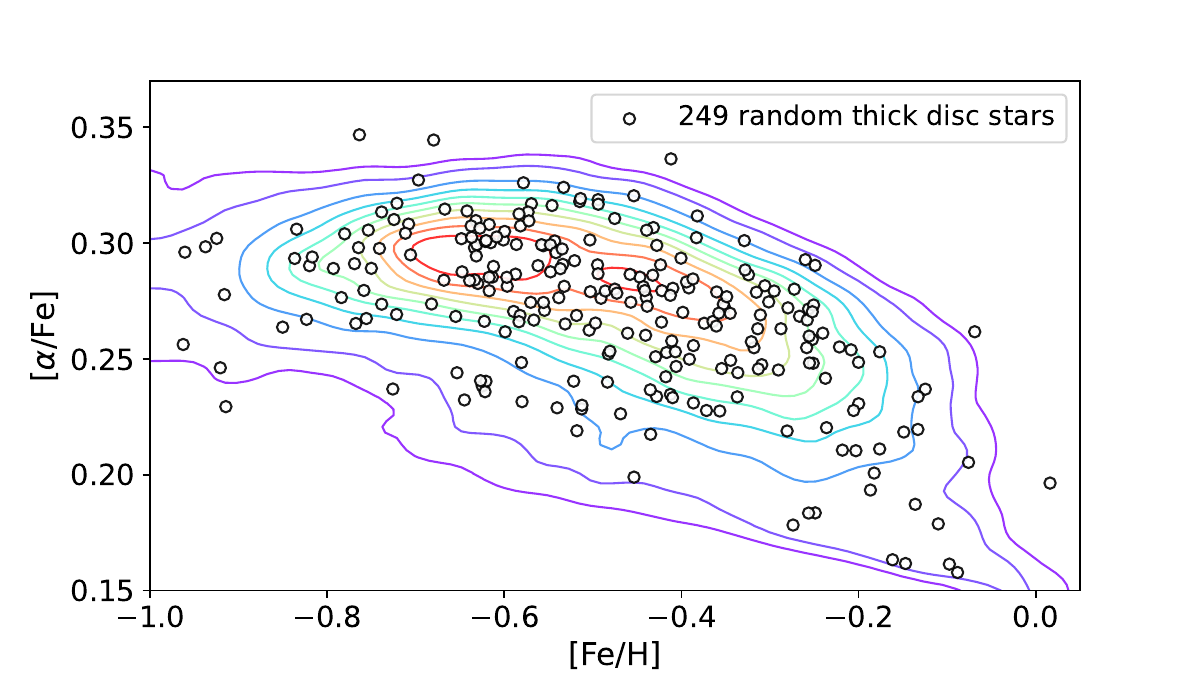}}
\caption{Chemical characterisation of the thick disc and YAR sample for red giant-type stars. \textit{Top panels}: MDFs and $\rm [\alpha/Fe]$ distribution functions of the YAR (in solid black lines) and thick disc (in dashed grey lines) samples. $\rm f = N_i/N_{tot}$ represents the fractional number density where $\rm N_i$ is the number of stars in each bin of $\rm [Fe/H]$ or $\rm [\alpha/Fe]$ and $\rm N_{tot}$ is the total number of stars of the sample.
\textit{Bottom panels}: $\rm [\alpha/Fe] - [Fe/H]$ distribution for YAR stars (black dots, left panel) and a randomly selected thick disc subsample (empty dots, right panel), both containing the same number of stars (written the relative legends.) The potential YAR binaries are denoted by squares: yellow squares indicate stars that exhibit at least one indicator of being a binary, while the magenta square represents the single star that fulfils all the examined indicators (see Section \ref{subsec: YAR-defin}). In both plots, the coloured lines represent stellar density iso-contours of the entire thick disc sample: 90$\rm \%$, 80$\rm \%$, 70$\rm \%$, 60$\rm \%$, 50$\rm \%$ , 40$\rm \%$, 30$\rm \%$, 20$\%$, 10$\rm \%,$ and 5$\rm \%$ of the peak density.} 
\label{chemics_RG}
\end{figure*}

In Fig.~\ref{fig: CN_thick_YAR}, we display the carbon and nitrogen distribution as function of the iron content for YAR stars and the thick disc sample. The plots in the first row show a noticeable stratification in age across the thick disc sample, with increasing age as $\rm [C/Fe]$ decreases and $\rm [N/Fe]$ increases.
This trend is observed for YAR objects as well, as emphasised in the bottom panels of Fig.~\ref{fig: CN_thick_YAR}, the YAR stars being located in the region of the plane where we expect younger (more massive) stars.
However we did notice two groups of outliers: YAR stars at $\rm [C/Fe] > 0.3$ (left panels of  Fig.~\ref{fig: CN_thick_YAR}) and  YAR stars at  $\rm [N/Fe] > 0.75$ (right panels of  Fig.~\ref{fig: CN_thick_YAR}). The latter group seem to be separated from the rest of YAR in the $\rm [N/Fe] - [Fe/H]$ plane, outlining a possible bi-modality in the nitrogen distribution. Moreover, these N\text-enhanced objects correspond to the youngest stars in the HRD of Fig.~\ref{fig: dhr} (left panel, dark blue colour). They occupy the leftmost position within the YAR sequence and do not overlap with the thick disc reference sample.
In the top-left panel of Fig.~\ref{chemics_RG}, we present the normalised histograms in metallicity and $\rm [\alpha/Fe]$ for the  YAR sample and the thick disc population in our dataset at $\rm R_{mean} < 20$ kpc. 
The two samples present some slight differences in the metal-poor tail of the distribution ($\rm -1 < [Fe/H] < -0.9$ dex) and at higher metallicities ($\rm -0.4 < [Fe/H] < -0.3$ dex), where the YAR distribution decreases sharply with respect to the thick disc.
Despite these dissimilarities, the YAR metallicity distribution function (MDF) generally resembles  the thick disc MDF, sharing similar characteristics and both peaking at $\rm [Fe/H] \sim -0.5$ dex. This result is confirmed by the findings of \citet{Sun2020} and \citet{Zhang2021}.
Similarly, the plot on the top right of Fig.~\ref{chemics_RG} highlights how the density histogram distribution in $\rm \alpha$\text-elements of the YAR sample clearly follows the thick disc star distribution. Indeed, they present similar shapes and matching peaks ($\rm [\alpha/Fe] \approx 0.28$ dex). However, we notice a small shift (possibly of  about a hundredth of dex) between the two. 
This shift is confirmed when looking at the distributions in the $\rm [\alpha/Fe] - [Fe/H]$ plane. The bottom-left plot of Fig.~\ref{chemics_RG} shows the stellar density iso-contours of the entire thick disc sample, together with the scatter distribution of the 249 YAR stars within 20 kpc. The yellow squares in the figure highlight the YAR stars showing at least one indicator of binarity, while the magenta square shows the only object that fulfils all indicator of binarity (see Section \ref{subsec: YAR-defin}). The points follow globally the thick disc distribution (including the stars targeted as binaries) but are shifted at lower $\rm [\alpha/Fe$] at any given metallicity.
For comparison, the bottom right plot of Fig.~\ref{chemics_RG} presents the analogous plot, this time showing the distribution of a sub-sample of thick disc stars randomly selected to contain the same number of stars of the YAR sample. The white points perfectly follow the thick disc contours.
We come back on this difference in the Section dedicated to detailed abundance ratios (Section \ref{sec:abundances}).


\subsection{Kinematics}\label{kinematics}

We now investigate the kinematics of the YAR stars in our dataset exploiting the kinematic and orbital parameters given in the astroNN catalogue. 
\begin{figure}
\resizebox{\hsize}{!}
{\includegraphics[width=10cm]{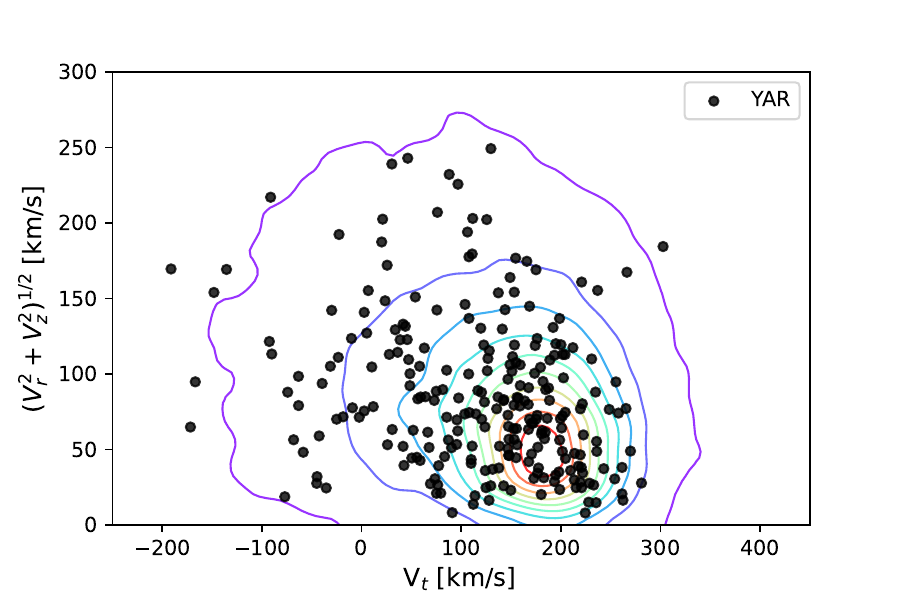}}
\resizebox{\hsize}{!}
{\includegraphics[width=10cm]{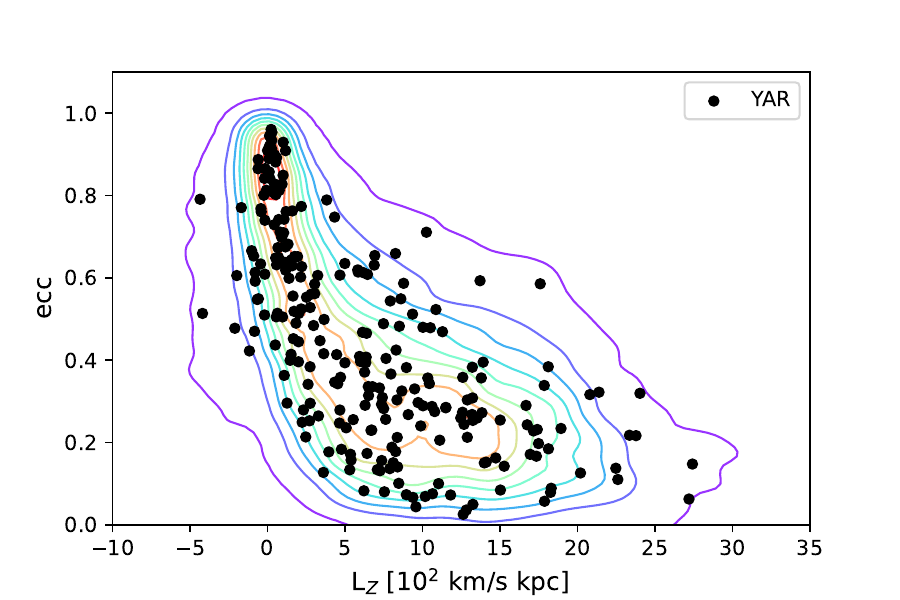}}
\resizebox{\hsize}{!}
{\includegraphics[width=10cm]{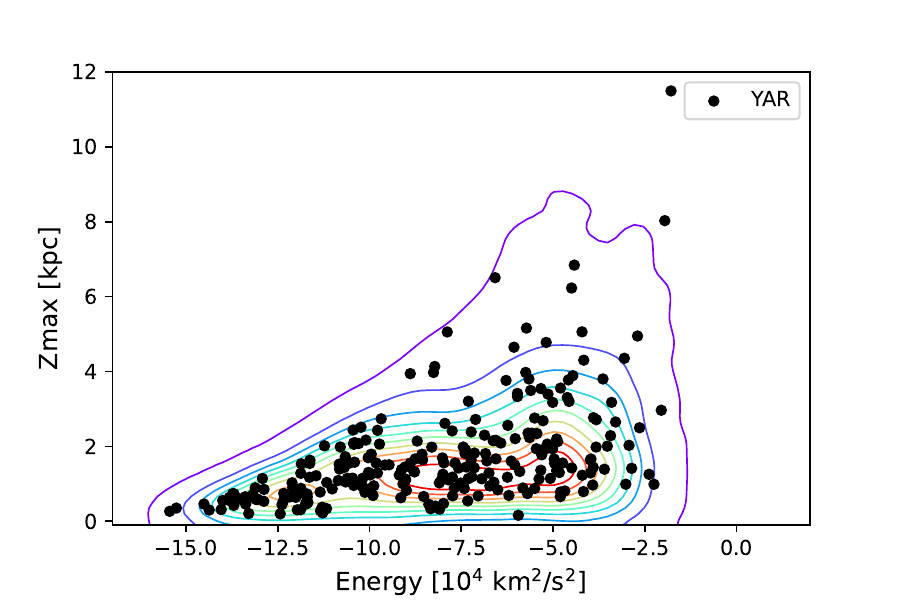}}
\caption{Dynamical properties of YAR stars (black dots) and thick disc sample (iso-density contours at 90$\rm \%$, 80$\rm \%$, 70$\rm \%$, 60$\rm \%$, 50$\rm \%$, 40$\rm \%$, 30$\rm \%$, 20$\rm \%$, 10$\rm \%,$ and 1$\rm \%$ of the peak density).}
\label{fig:Toomre_contour}
\end{figure} 
Figure~\ref{fig:Toomre_contour} summarises these properties of our YAR sample compared to stars of the thick disc.
YAR stars present a large spread in the Toomre diagram (top panel) resembling the characteristics of thick disc stars, the distribution of which is represented by iso-density contours in the figure. As shown by the plot, these objects span a considerably large range of $\rm V_{r}$ and $\rm V_{z}$ velocities and tend to rotate more slowly than the younger disc.
The middle panel of Fig.~\ref{fig:Toomre_contour} represent another useful kinematic diagnostic: the eccentricity - angular momentum L$_{z}$ plane. thick disc stars tend to distribute along the entire range of eccentricity and the majority of them present low values of $\rm L_{z}$. YAR stars follow this distribution perfectly. The tail at high eccentricities is due to stars that are trapped in the bar.
Finally, in the bottom panel of Fig.~\ref{fig:Toomre_contour} , we show the Zmax (maximum height reached from the Galactic plane) -- E (energy) distribution. Once again, the kinematic properties of YAR stars match those of the thick disc.
In conclusion, the overall kinematic properties of YAR stars are consistent with the thick disc. This result is in agreement with the findings of previous works \citep{Zhang2021, Jofre2022}.

\section{Accreted stars}
\label{sec:accreted}
\begin{figure*}
\resizebox{\hsize}{!}{
\subfloat[][Chemical plane]{\includegraphics[width=10cm]{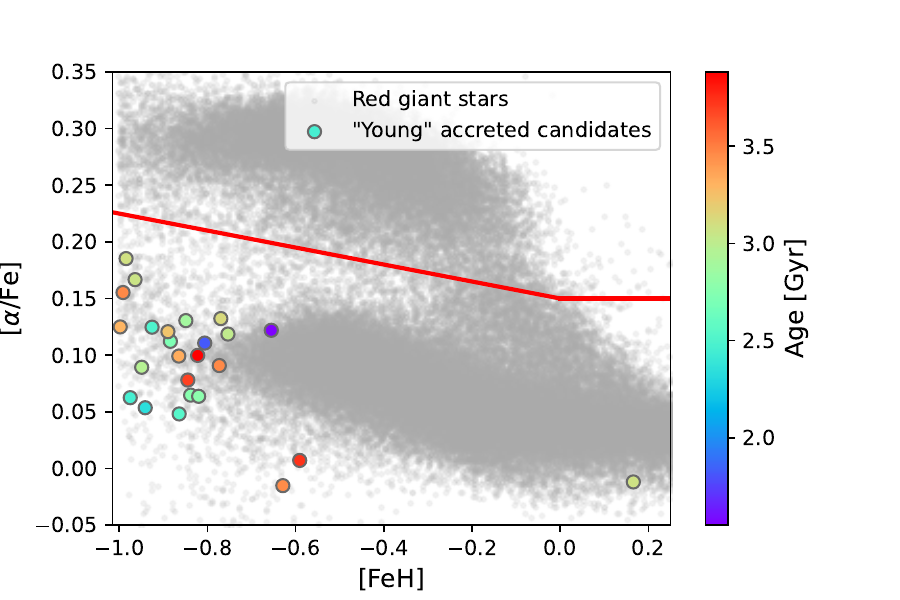}} 
\subfloat[][Toomre diagram]{\includegraphics[width=10cm]{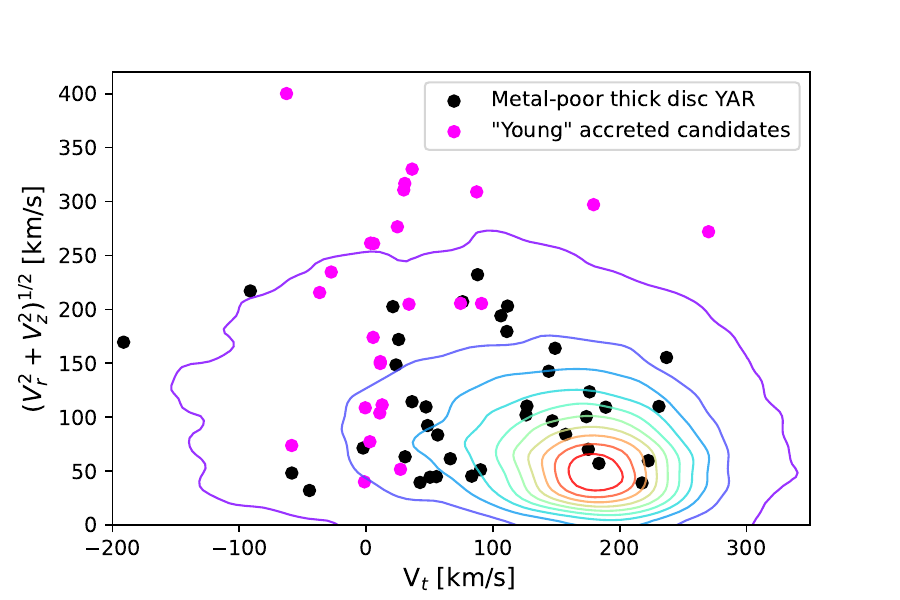}}}
\resizebox{\hsize}{!}{
\subfloat[][HRD]
{\includegraphics[width=10cm]{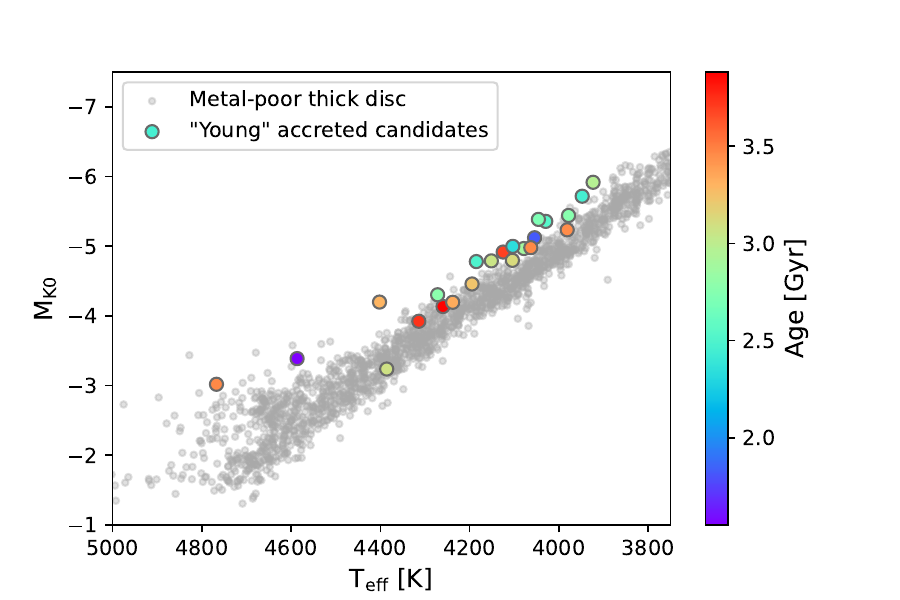}}
\subfloat[][Mass distributions]{\includegraphics[width=10cm]{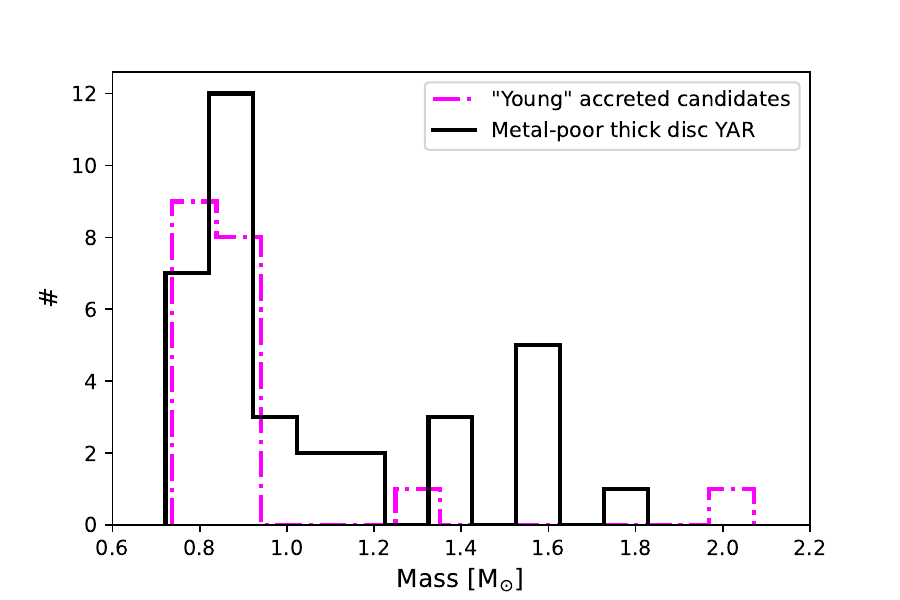}}}
\caption{Chemo-dynamical characterisation of the young accreted candidate sample selected having eccentricity $\rm >$ 0.7 and $\rm R_{APO} > 10$ kpc (see the text for details). The sample is colour-coded by age in \textit{panels (a) and (c)} and is shown in fuchsia in \textit{panels (b) and (d)}. YAR giants stars having $\rm [Fe/H] < -0.8$ dex are represented in black in the \textit{panels (b) and (d)}. The under-plotted distribution in the $\rm [\alpha/Fe]-[Fe/H]$ plane represents the red giants reference population in APOGEE DR17, while in the HRD, the underlying comparison sample is made of thick disc stars stars having $\rm [Fe/H] < -0.8$ dex.}
\label{fig:accreted}
\end{figure*}
As introduced in Section \ref{sec:intro}, the BSSs are the product of a mass acquisition event, either via mass transfer or via merger. Considering that these phenomena are known to occur in other galaxies \citep[e.g.][]{Momany2015} we now go on to look at stragglers candidates in the accreted stars in APOGEE DR17. The selection of halo-like population that we describe here was done with an awareness of the complexity of defining a clear and uncontaminated sample of likely accreted stars.
We chose red giants with an eccentricity greater than 0.7, orbital apocentre $\rm (R_{apo})$ greater than 10 kpc, gravitationally bound to the Galaxy (E$\rm _{orb} < 0\,  km^{2}/s^{2}$), and with errors on distances less than 1.5 kpc, similarly to what was done in, for instance, \citet{Myeong2022}.
In order to focus on the most likely accreted objects, we additionally removed from the defined halo sample the heated thick disc population (metal rich halo-like component interpreted as the in situ constituent of the Galactic halo consisting of kinematically heated thick disc stars \citep{DiMatteo2019b, Belokurov2020}, picking only stars at $\rm [\alpha/Fe] < -0.075 \times [Fe/H] + 0.15$ dex. We obtain 254 stars, 25 of which are young (age $\rm<$ 4 Gyr and error on age $\rm <$ 3 Gyr).
Their position in the $\rm [\alpha/Fe]-[Fe/H]$ plane is highlighted in the panel (a) of Fig.~\ref{fig:accreted}. Most of these stars are characterised by metallicity below $\rm -$0.6 $\rm /$ $\rm -$0.8 dex\footnote{We remark that in general we discard from the APOGEE DR17 sample the stars having $\rm [Fe/H] < -1$, the age estimate of them being not reliable \citep{Mackereth2019}.} and $\rm [\alpha/Fe] < 0.20$ dex.
We show the distribution of the young accreted candidates in the HRD of panel c in Fig.~\ref{fig:accreted}. They form a sequence located to the left of the thick disc stars in the same range of metallicity ([Fe/H] $\rm <$ -0.8 dex), which is as expected if they are indeed more massive. 
The distribution in the Toomre diagram (plot b) illustrates the difference in kinematics between accreted YAR stars and those of the thick disc over the same metallicity interval. Most accreted YAR stars have galactic rotational velocity below 50 km $\rm s^{-1}$, while most thick disc stars are above this limit, and in agreement with the main thick disc population.
Panel d of Fig.~\ref{fig:accreted} shows the mass distributions of the young accreted sample compared to the YAR stars in the same range of metallicity ($\rm [Fe/H] < -0.8$ dex). The mass determination is available for 19 stars and 35 stars, respectively.
The distributions are not normalised and the y-axis represents the real number of stars in each bin of mass. The mass distributions of the two populations slightly differ from each other.
This is contrary to the expectations, given the position of the young accreted stars on the HRD. 
This sample has indeed masses which peak at around 0.9 $\rm M_{\odot}$. The distribution does not extend to the higher end of the mass range, except for 2 single stars (around 1.3 $\rm M_{\odot}$ and 2.0 $\rm M_{\odot}$).
It is possible that our sample is not extensive enough and that we do not dispose of the statistics necessary to study this population of stars, especially its mass distribution.
In addition, it is plausible that
the $\rm [C/N$] ratio exploited to determine the ages listed in the astroNN catalogue (following the method highlighted in \citet{Mackereth2019}) could lose sensitivity at low metallicities, causing a less reliable mass (and consequently age) estimate.

To conclude, given the observational evidence obtained with this current dataset, we cannot claim with confidence that the accreted halo-like population contains rejuvenated objects.

\section{Other abundance ratios}\label{sec:abundances}

Chemical abundances can provide additional interesting clues on the possible formation pathways of YAR stars.
Indeed, the formation pathways of stragglers lead to different chemical signatures, depending on how the mass was acquired (mass transfer or collision) and on the evolutionary phase of the donor star at the time of the straggler formation. \citet{Lombardi1995, Lombardi1996} and \citet{Sills2001} predict that the collision product will retain a chemical profile very similar to that of the parent stars, and so little effect on the surface abundances is expected. On the contrary, if stragglers are formed by mass transfer from an evolved star to its companion,  surface contamination by elements typical of the advanced stage of stellar evolution (s-neutron capture elements) is expected \citep[i.e.][]{Boffin1988}. Below, we check whether the chemical abundances of our sample of YAR stars differ from those of the parent thick disc population.

 
\subsection{$\alpha$\text-elements}
\begin{figure*}
\resizebox{\hsize}{!}{
\includegraphics[width=9.5cm]{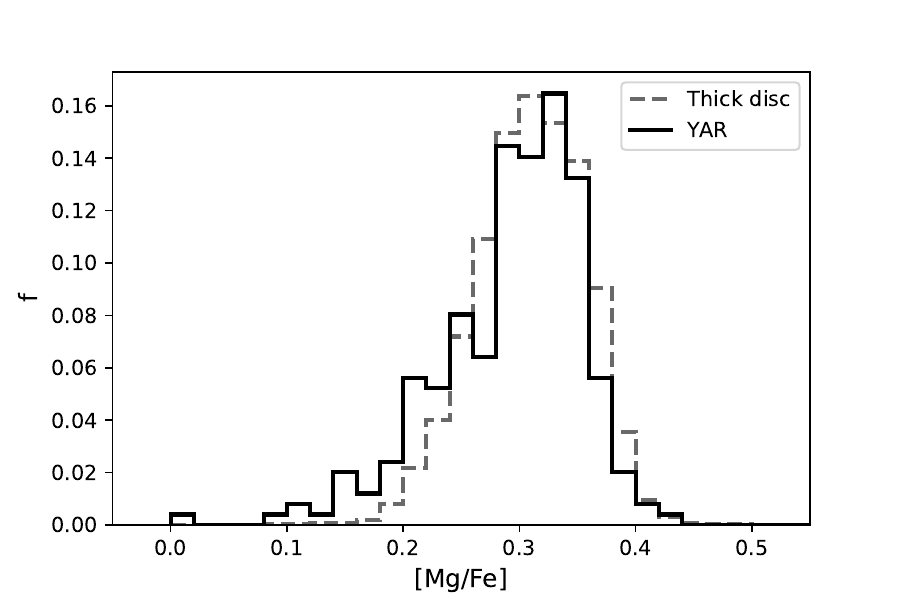}
\includegraphics[width=9.5cm]{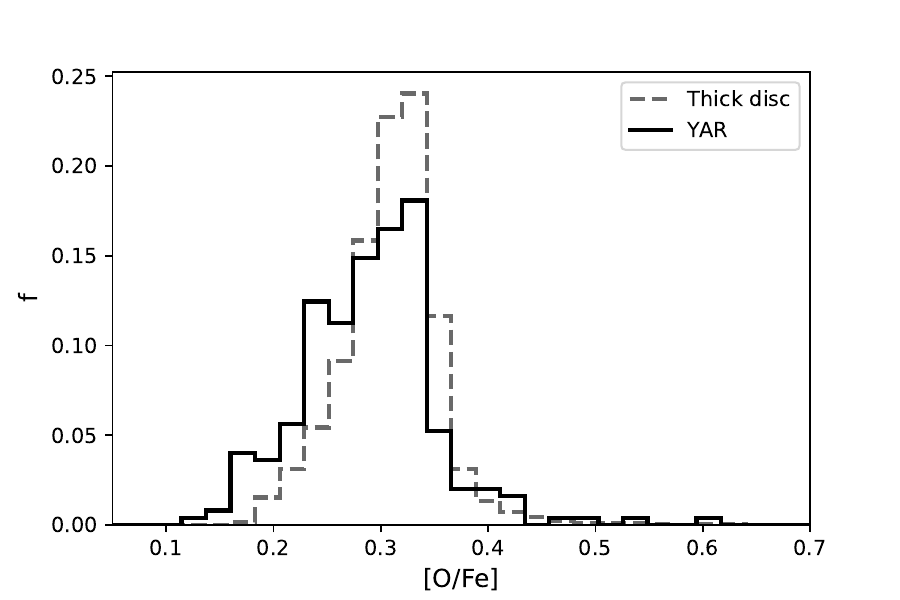}}
\resizebox{\hsize}{!}{
\includegraphics[width=9.5cm]{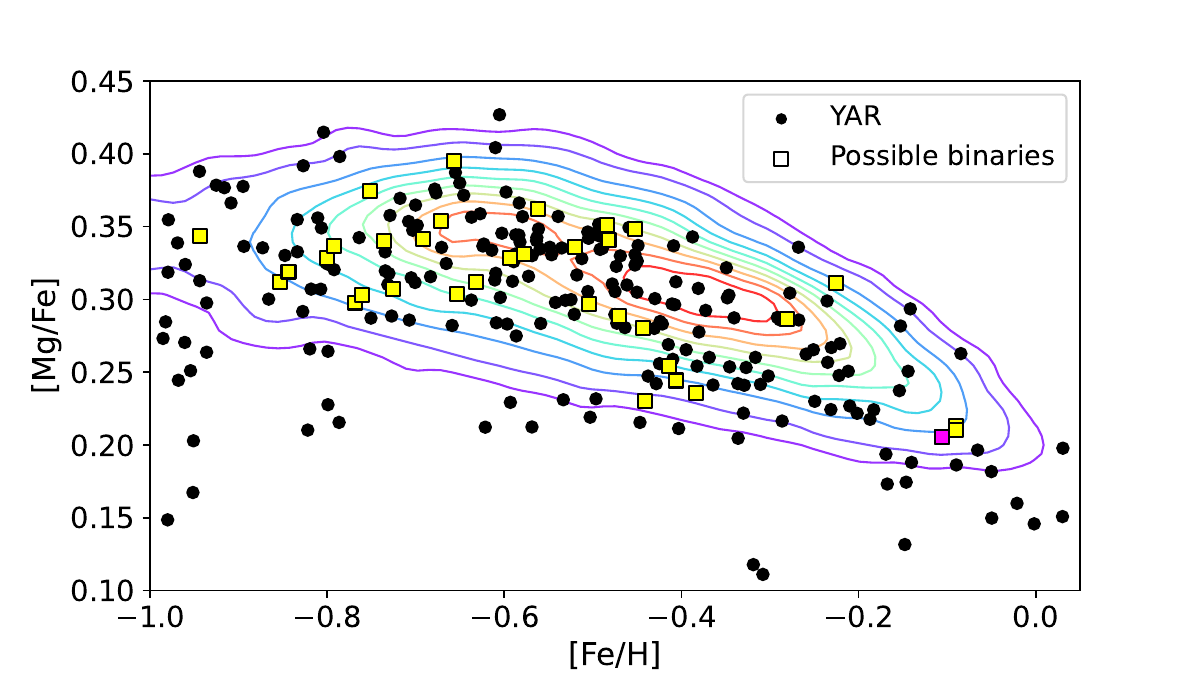}
\includegraphics[width=9.5cm]{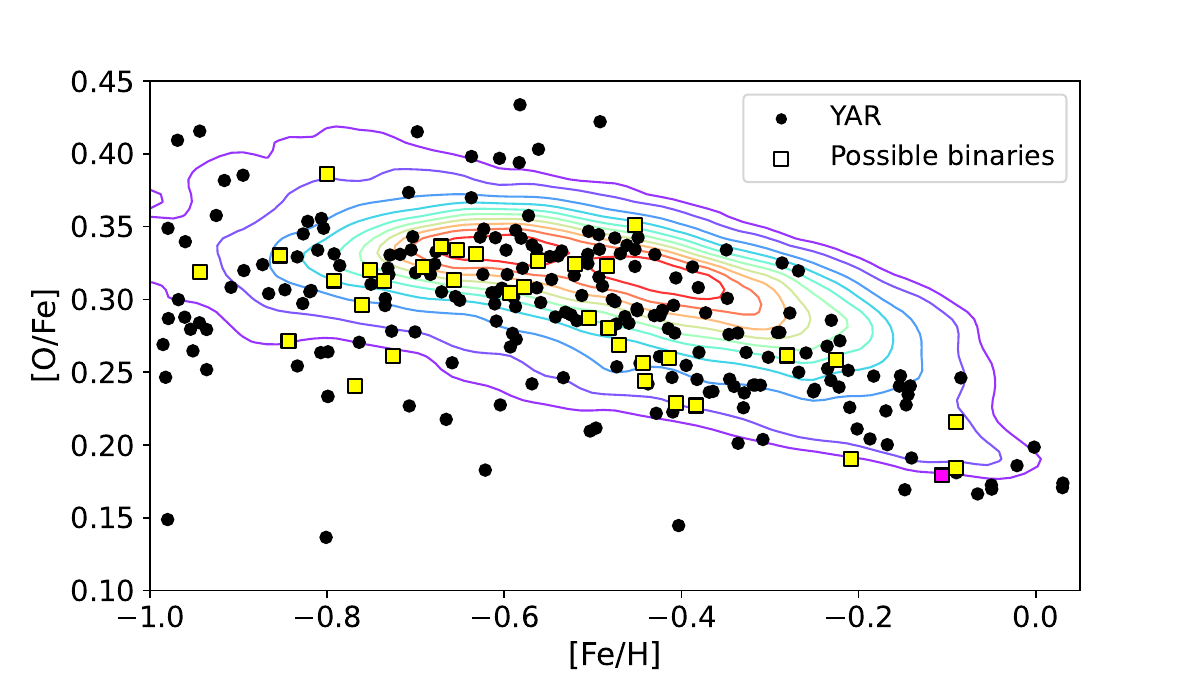}}
\caption{Comparison of magnesium and oxygen abundance for stars in the YAR sample and stars in the thick disc. \textit{Top panels}: Thick disc sample is represented by the grey dashed histogram, while the distribution of YAR stars is represented by the black solid histogram. \textit{Bottom panels}: Thick disc reference sample is highlighted with iso-density contours at 90$\%$, 80$\%$, 70$\%$, 60$\%$, 50$\%$, 40$\%$, 30$\%$, 20$\%$, 10$\%,$ and 5$\%$ of the peak density, while the YAR sample is represented by black dots. The potential YAR binaries are denoted by squares: yellow squares indicate stars that exhibit at least one indicator of being a binary, while the magenta square represents the single star that fulfils all the examined indicators (see Section \ref{subsec: YAR-defin}).}
\label{fig:Mg-O}
\end{figure*}
\begin{figure*}
\resizebox{\hsize}{!}{
\includegraphics[width=9.5cm]{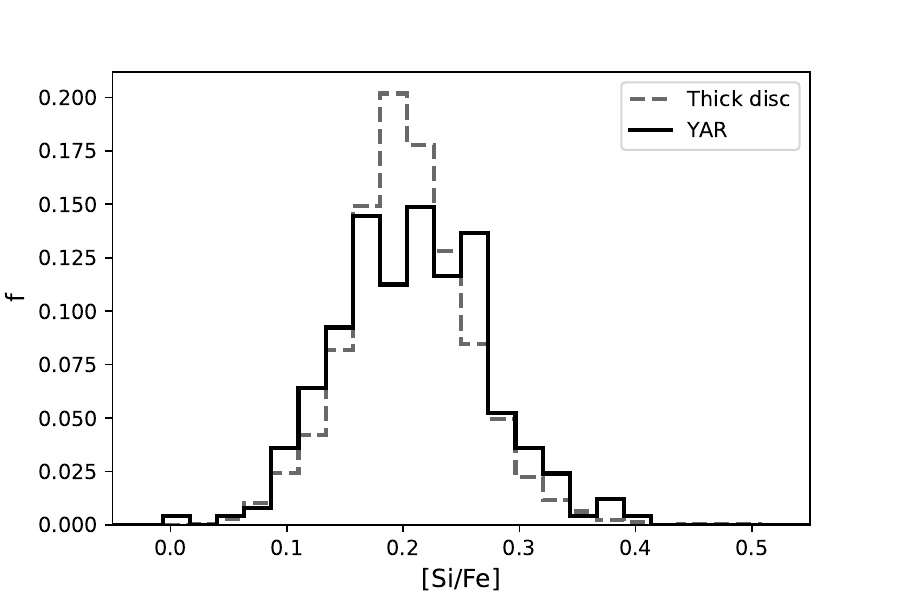}
\includegraphics[width=9.5cm]{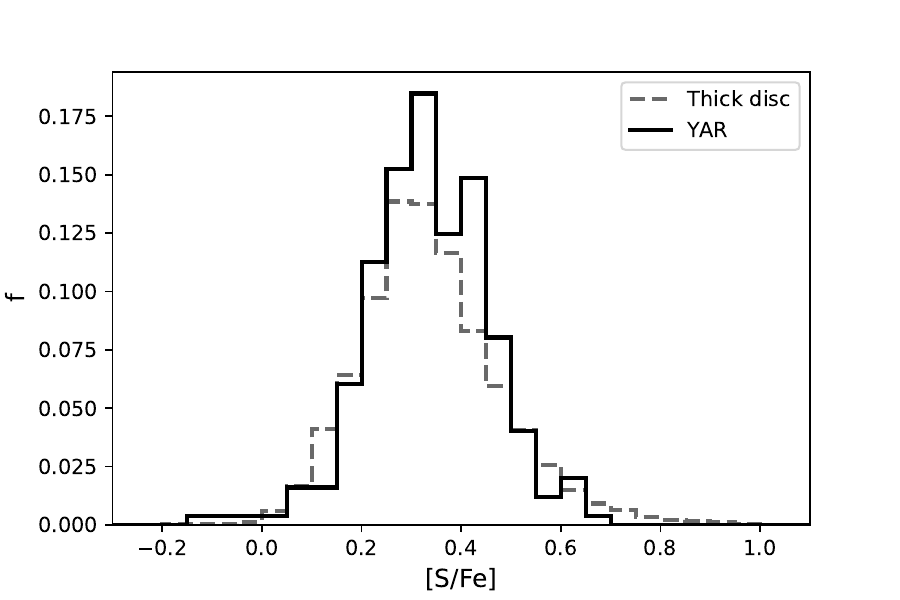}}
\resizebox{\hsize}{!}{
\includegraphics[width=9.5cm]{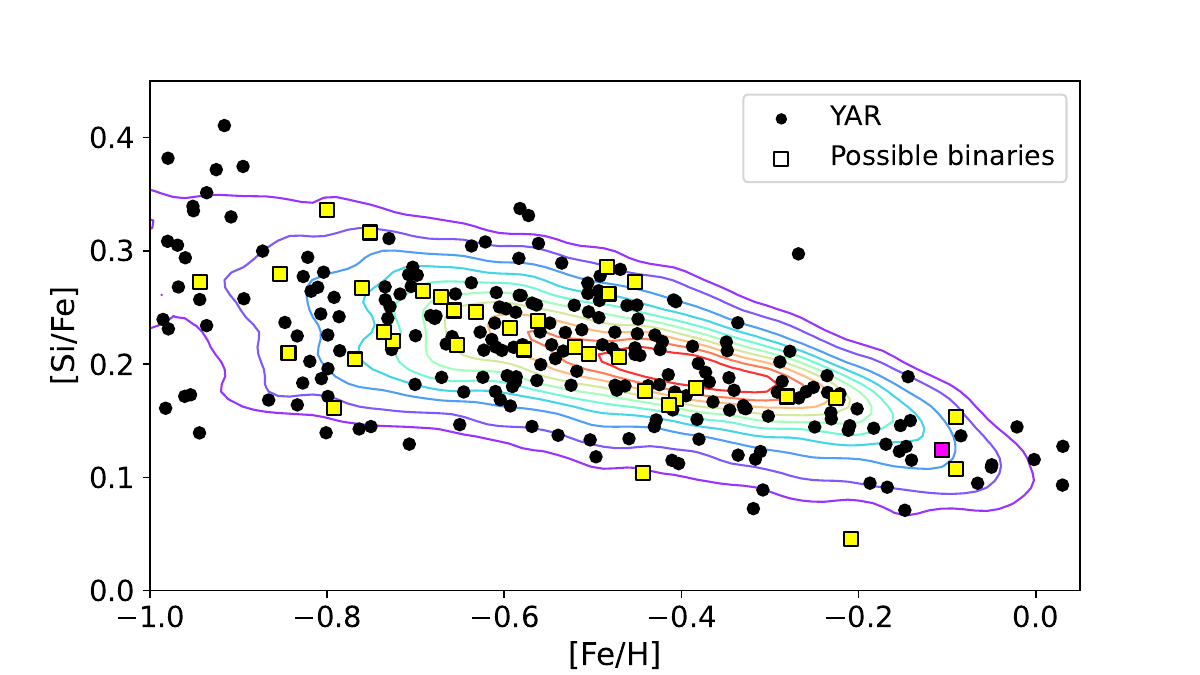}
\includegraphics[width=9.5cm]{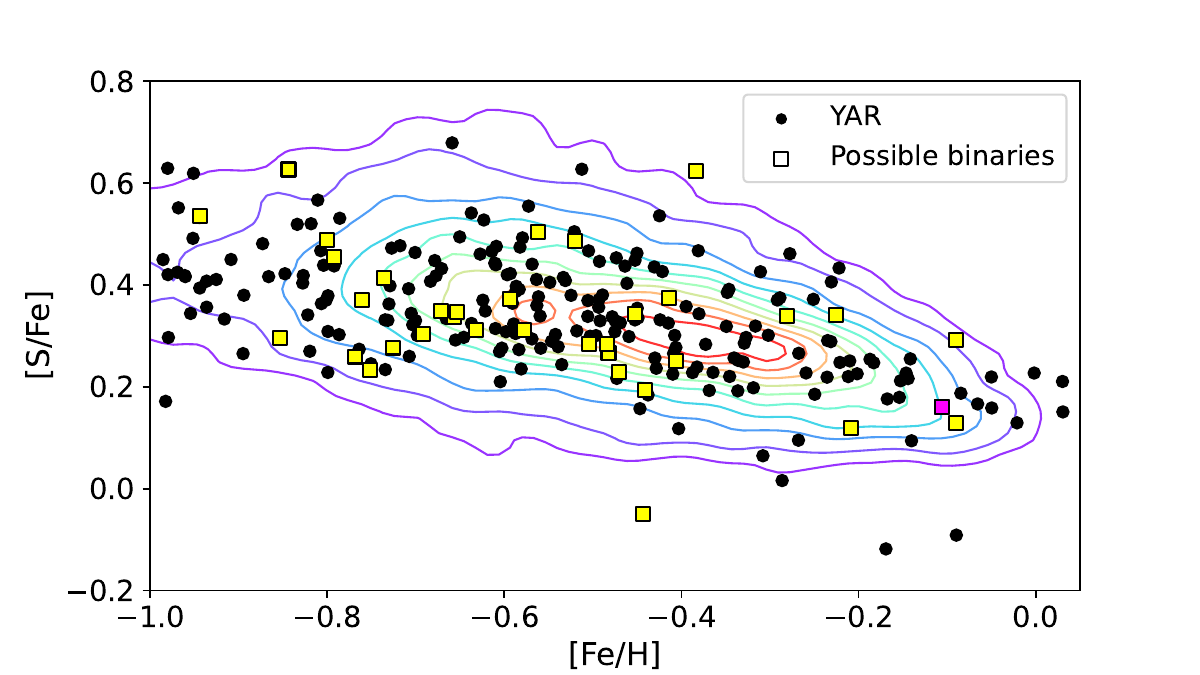}}
\caption{Same as Fig.~\ref{fig:Mg-O} for Si and S.}
\label{fig: Si-S}

\end{figure*}
\begin{figure*}
\resizebox{\hsize}{!}{
\includegraphics[width=9.5cm]{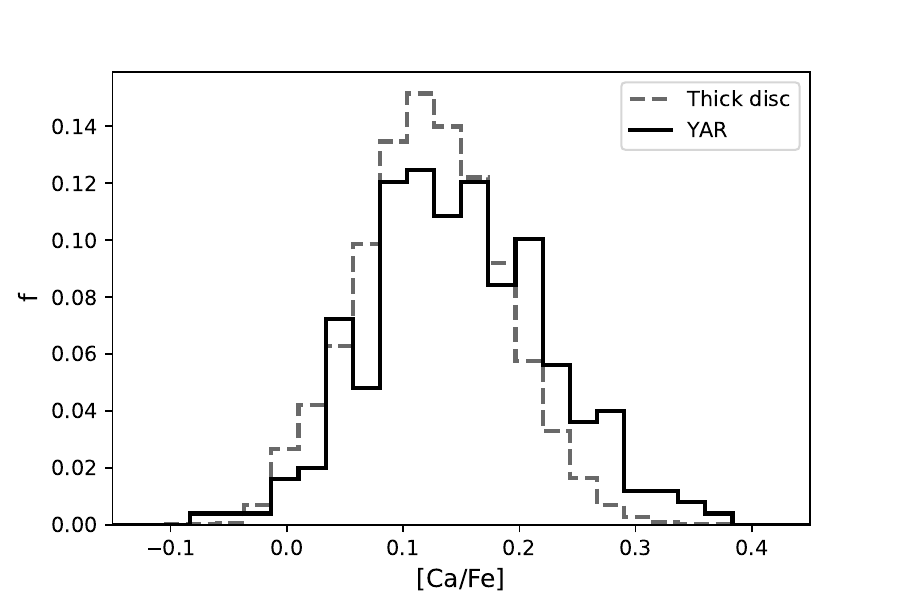}
\includegraphics[width=9.5cm]{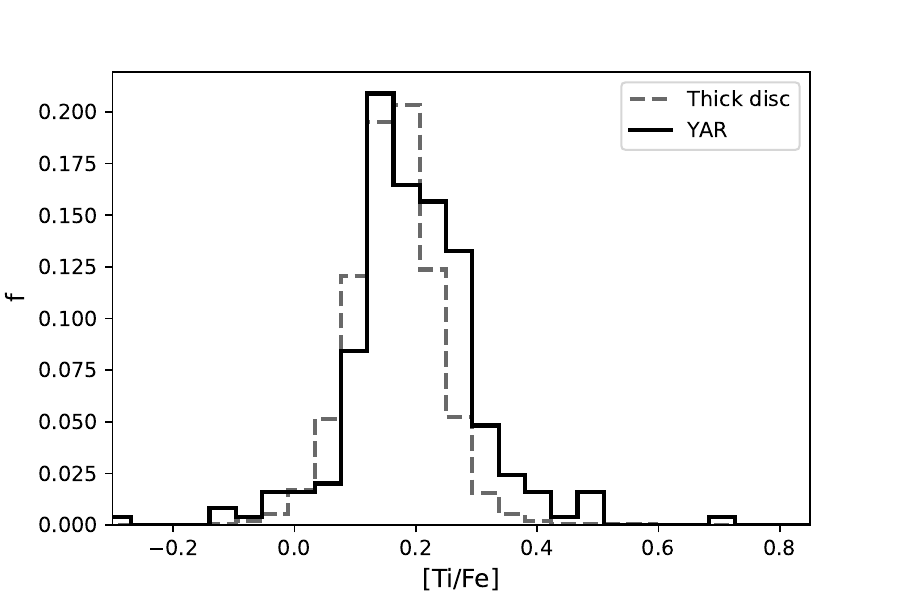}}
\resizebox{\hsize}{!}{
\includegraphics[width=9.5cm]{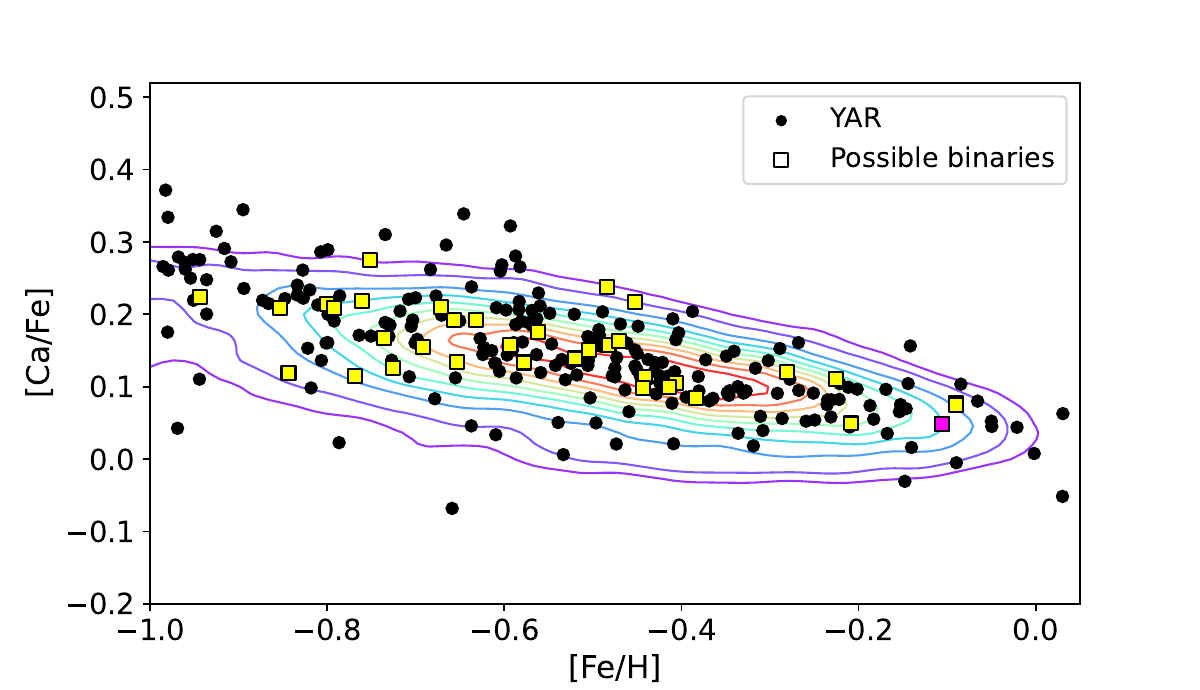}
\includegraphics[width=9.5cm]{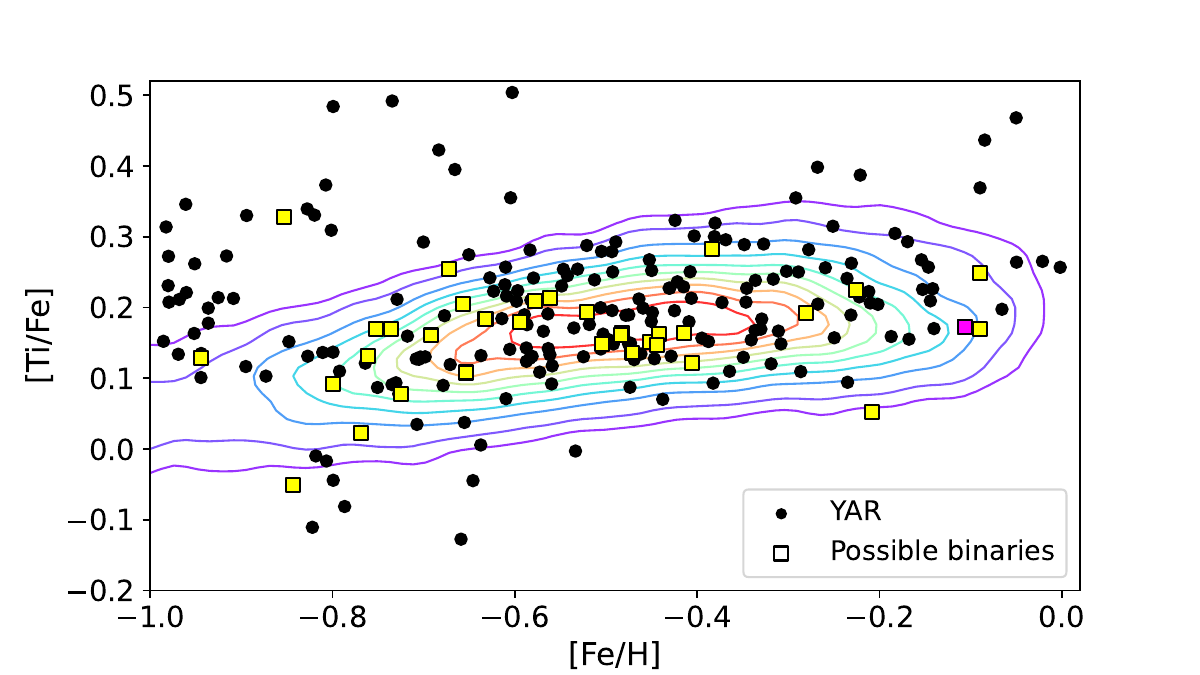}}
\caption{Same as Fig.~\ref{fig:Mg-O} for Ca and Ti.}
\label{fig: Ca-Ti}
\end{figure*}

In the APOGEE DR17 survey, the $\rm \alpha$\text-elements abundance is defined as the combination of O, Mg, Si, S, Ca, and Ti. To investigate which of these chemical elements are driving the trend of the YAR distribution described in Section \ref{sub: metallicity-alpha} over the others, we examined the $\rm [X/Fe]-[Fe/H]$ distributions, in addition to the normalised histograms, for each individual $\rm \alpha$\text-element. We note that 246 stars and 231 stars out of 249 present estimates for S and Ti, respectively (while the other $\rm \alpha$\text-element abundances are listed for the entire YAR sample).
We found that Mg and O seem to be the main cause of the discrepancy in the $\rm [\alpha/Fe]-[Fe/H]$ global distribution. Figure~\ref{fig:Mg-O} shows the tendency of the YAR sample distributions to be shifted to lower values with respect to the thick disc sample's distributions, particularly at $\rm [Fe/H] > -0.4$ dex. This trend is confirmed by the measurements of O and Mg from the astroNN catalogue, which show an even larger discrepancy.
Si and S, on the contrary, do not display discrepancies between the two populations, as shown in Fig.~\ref{fig: Si-S} (this is also confirmed by the astroNN abundances).
From Fig.~\ref{fig: Ca-Ti} it is noticeable that Ca and Ti distributions exhibit a noticeable shift between the two samples, opposite to the trend for oxygen and magnesium. However, these differences are not confirmed by the astroNN abundances. In particular, the distributions of YAR and thick disc stars in Ca obtained from astroNN show no difference, while the Ti distributions exhibit a trend opposite to the APOGEE Ti abundance determination ($\rm Ti_{YAR} < Ti_{THICK\, DISC}$). The stars that were targeted as possible binaries, indicated by yellow and magenta squares in the corresponding figures, conform to the overall YAR trend for each $\rm \alpha$\text-element examined.
Our results are partially confirmed by \citet{Jofre2022} who found that their sample of 'over-massive' stars follow the trend in the individual $\alpha$\text-capture elements (from APOGEE DR16) of the thick disc, while the 'under-massive' stars show slightly lower abundances.
An additional validation of our results comes from \citet{Zhang2021}, whose Fig. 8 shows shifts in $\alpha$\text-elements distribution comparable to what we find with APOGEE DR17. Only the titanium trend in \citet{Zhang2021} presents an inconsistency with APOGEE, displaying a shift more compatible with the astroNN trend. We remark, however, that titanium has been noted as problematic in APOGEE DR17 documentation\footnote{\url{https://www.sdss4.org/dr17/irspec/abundances/}} for giants stars,  with its measurement in APOGEE DR17 deviating from the literature expectations. 
For this reason, we tend to give more credence to astroNN and \citet{Zhang2021} titanium measurements.
We note, however, that even in these cases the difference in the distribution of the two populations is small ($\sim$0.02 dex) and possibly not significant.

\subsection{Al \& Na}
\begin{figure*}
\resizebox{\hsize}{!}{
\includegraphics[width=9.5cm]{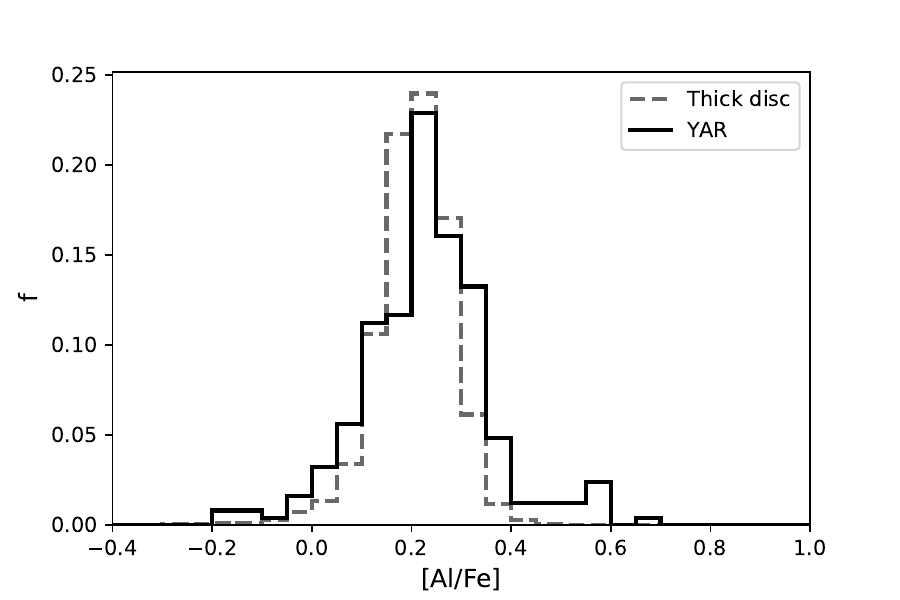}
\includegraphics[width=9.5cm]{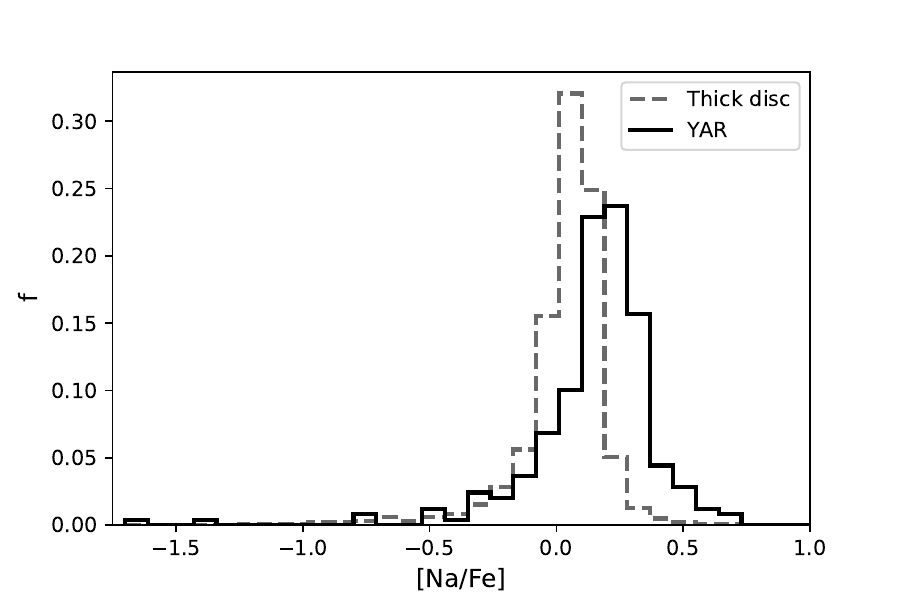}}
\resizebox{\hsize}{!}{
\includegraphics[width=9.5cm]{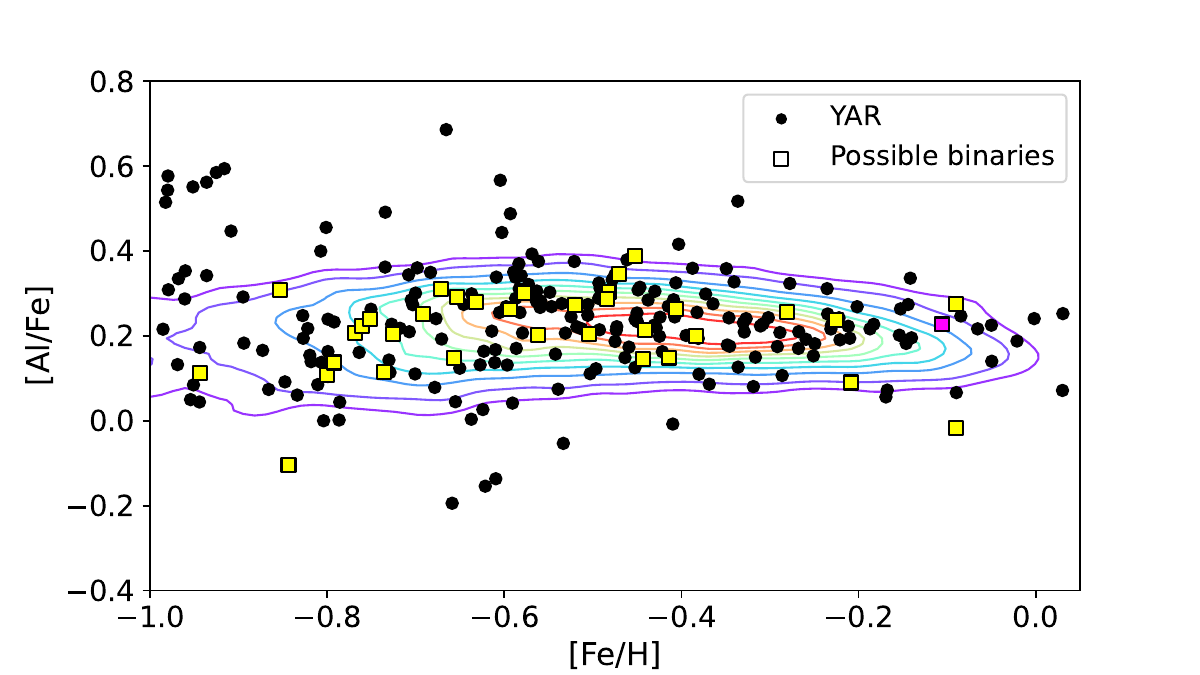}
\includegraphics[width=9.5cm]{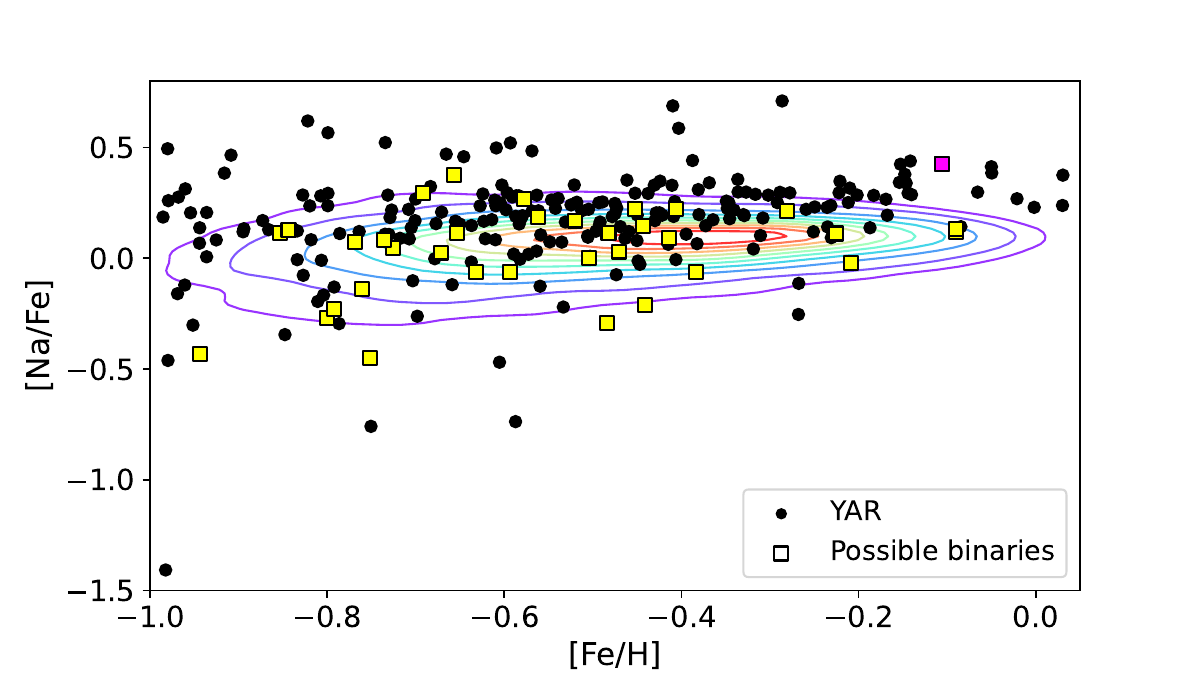}}
\caption{Same as Fig.~\ref{fig:Mg-O} for Al and Na.} 
\label{fig:Al}
\end{figure*}
\begin{figure}
\includegraphics[width=\hsize]{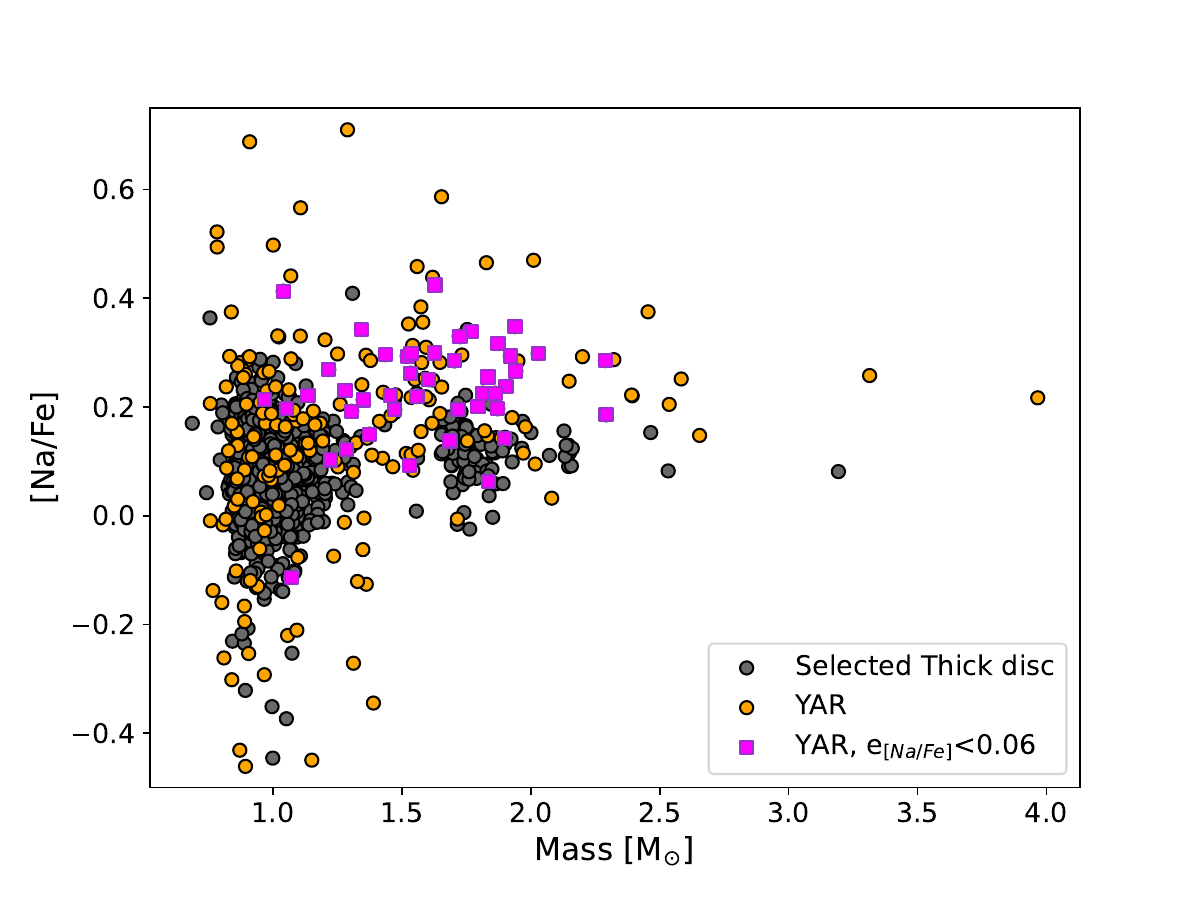}
\caption{$\rm [Na/Fe] - mass$ distribution of our samples. The orange dots represent the YAR stars; the fuchsia squares show the YAR stars with an uncertainty on $\rm [Na/Fe] < 0.06$ dex; the grey dots symbolise the thick disc stars, selected to have an uncertainty on $\rm mass   < 0.05\, M_{\odot}$ and uncertainty on Na abundance $\rm <$ 0.06 dex.}
\label{fig:Na}
\end{figure}
Overall, 248 stars and 246 stars out of 249 present estimates for Na and Al, respectively.
Figure~\ref{fig:Al} shows that the YAR stars are characterised by a content in Al that is consistent with the thick disc reference sample. This is in agreement with astroNN catalogue, as well as the findings of \citet{Zhang2021} and \citet{Jofre2022}. 

The plots of Fig.~\ref{fig:Al} show a  difference of about $\rm \sim$ 0.1 dex in sodium abundance between the YAR stars and thick disc sample. 
The Na content is shifted towards higher values, although this variation is not confirmed by the astroNN data. 
\cite{Zhang2021} discarded the Na abundances from the LAMOST spectra 
in their analysis because of the potential contamination of sodium features by the interstellar medium.
On the other hand, \citet{Jofre2022}  stressed a scattered distribution with different YAR stars falling off the overall disk distribution. This is justified by the authors remarking that Na abundance in APOGEE DR16 was not derived with particularly high precision, having been obtained from two weak lines \citep{Jonsson2020}.
The BAWLAS catalogue from \citet{Hayes2022} provides an additional term of comparison for problematic chemical elements for about 120,000 giant stars in APOGEE DR17. 
The check done with BAWLAS confirms the presence of a shift between the YAR stars and the thick disc distributions of about 0.15$/$0.20 dex in Na, suggesting that the shift between YAR and the standard thick disc population is real. 

It has been suggested that mixing could modify the sodium abundance depending on the mass of the star \citep[see][]{Luck1994, Smiljanic2009, Smiljanic2012}.
We plot in Fig.~\ref{fig:Na} the Na abundance as a function of stellar mass, for the sample with available mass determinations (see Section \ref{subsec:dhr}). The standard thick disc sample is further reduced by selecting stars with the best $\rm [Na/Fe]$ abundance (error on $\rm [Na/Fe]$ smaller than 0.06 dex) and presented with grey dots. The YAR stars are plotted as orange dots, while those with an error smaller than 0.06 dex on sodium abundance are highlighted with a fuchsia square symbol. The larger spread in sodium abundance at masses less than 1.2 $\rm M_{\odot}$ is an effect of the larger uncertainties in this mass range.
The figure shows that thick disc stars are mostly concentrated in mass below 1 $\rm M_{\odot}$, with a small group around 1.7 $\rm M_{\odot}$. These peculiar thick disc stars exhibit lower sodium dispersion compared to the remaining thick disc stars at lower masses. By considering their position in the HRD, we observe that they are more luminous than the typical thick disc objects. However, the exact reason behind their increased mass remains uncertain.
We see that YAR stars are much more dispersed in term of mass, with about $\rm 54 \%$ of the sample having a mass higher than 1.2 M$_{\odot}$. The plot shows that the shift to higher $\rm [Na/Fe]$ values is mainly due to these stars: the mean value of sodium content above this limit is in fact $\rm [Na/Fe] = 0.21$ dex, while it is $\rm [Na/Fe] = 0.06$ dex below.
The lack of YAR stars at $\rm [Na/Fe] < 0$ dex and masses greater than 1.5 M$_{\odot}$ suggests that the enhanced sodium abundance is linked to the mass of stars. 
Stellar models do not uniformly predict an increase of sodium abundance with stellar mass in the mass range of our sample (1\text-2 $\rm M_{\odot}$). 
For example, models with rotation from \citet{Lagarde2012} show a steep increase between about 1 $\rm M_{\odot}$ and 2.4 $\rm M_{\odot}$, while models without rotation remain flat to about 2 $\rm M_{\odot}$.

\subsection{Ce \& V}
\begin{figure*}
\resizebox{\hsize}{!}{
\includegraphics[width=9.5cm]{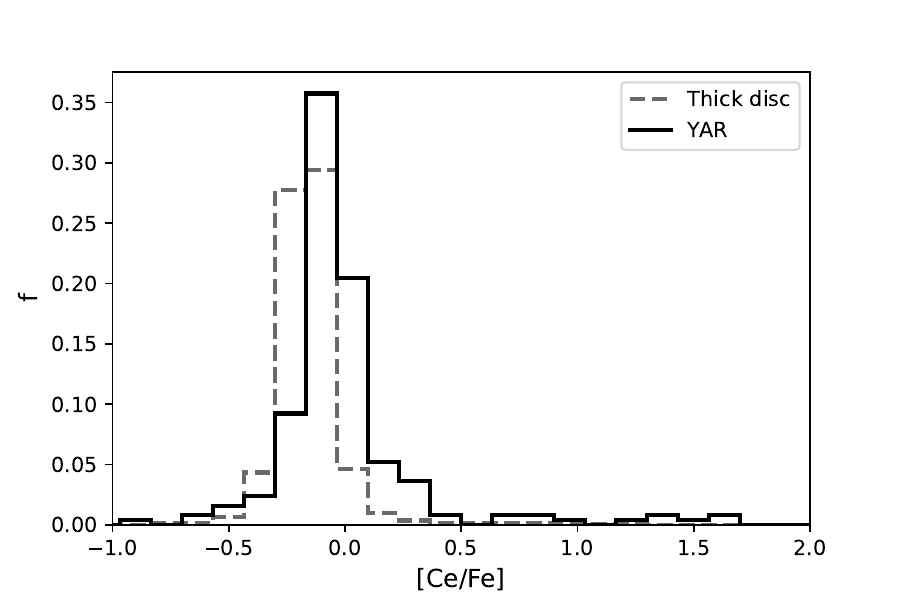}
\includegraphics[width=9.5cm]{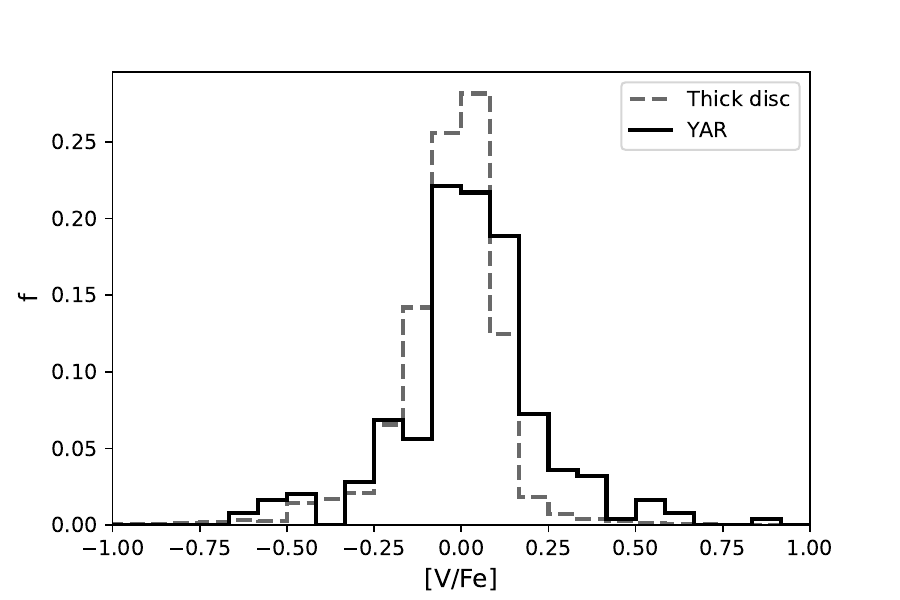}}
\resizebox{\hsize}{!}{
\includegraphics[width=9.5cm]{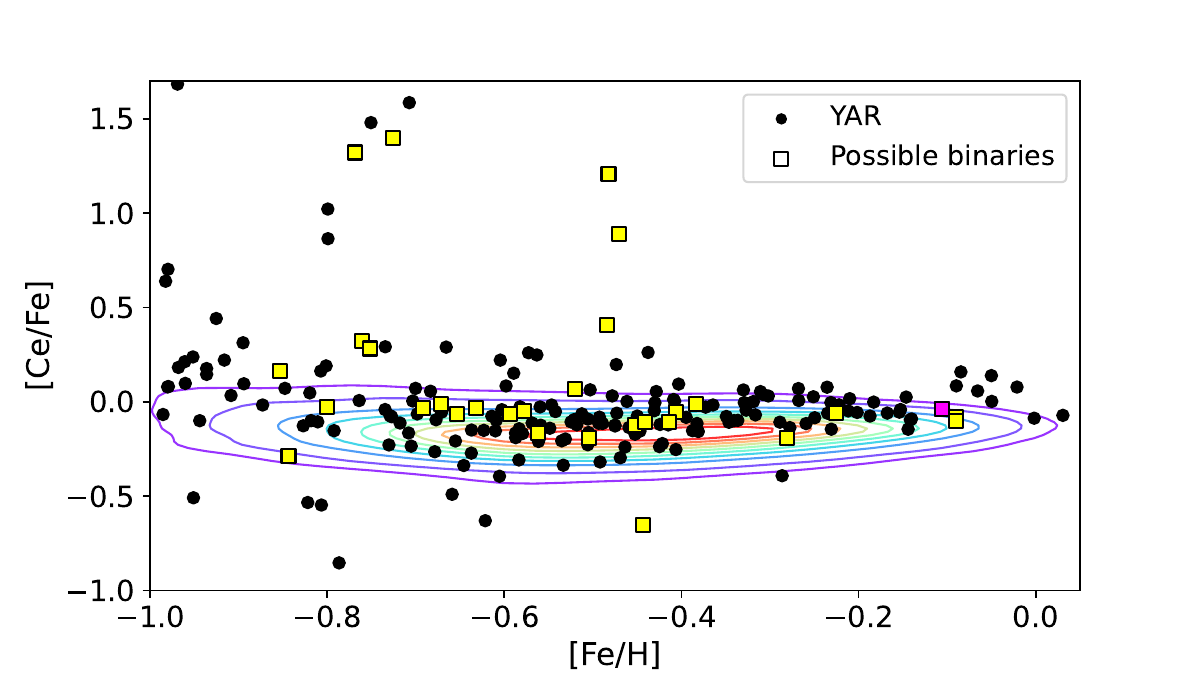}
\includegraphics[width=9.5cm]{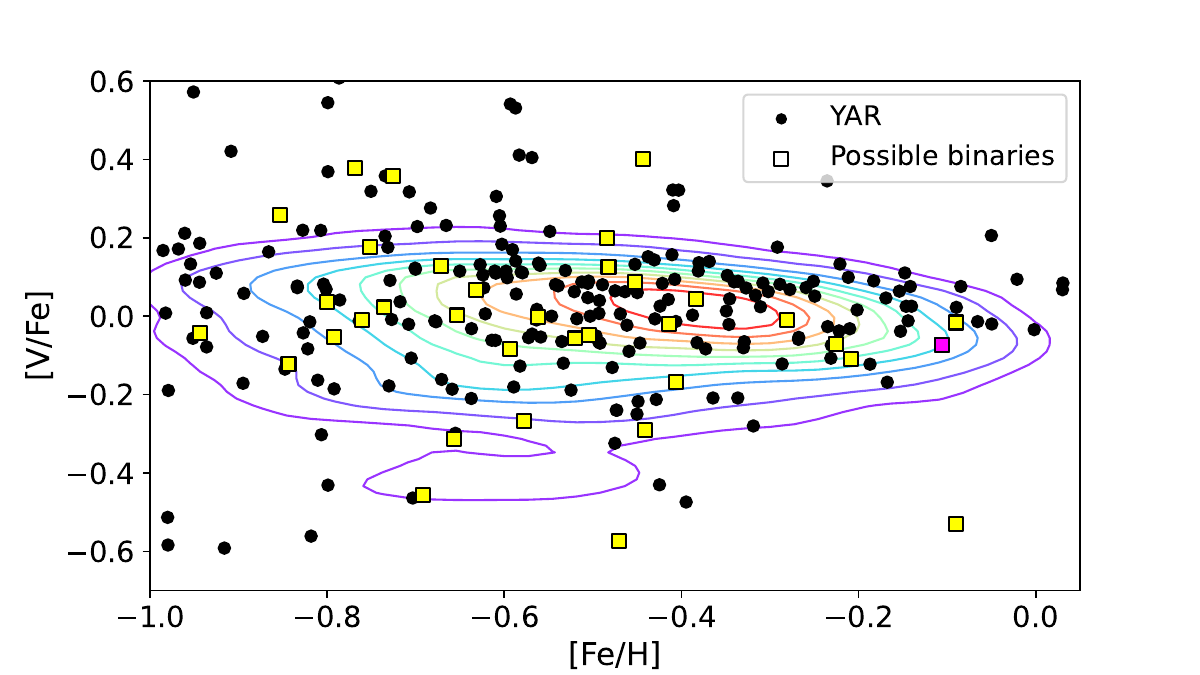}}
\caption{Same as Fig.~\ref{fig:Mg-O} for Ce and V} 
\label{fig:Ce-V}
\end{figure*}
In Fig.~\ref{fig:Ce-V}, we report our results on cerium and vanadium.
We note that 211 stars stars out of 249 have listed estimates of Ce (while every star in the YAR sample has a V estimate).
The distributions show that YAR objects are more Ce enhanced with respect to the thick disc stars. This is confirmed by the BAWLAS abundances, with a similar offset, although the absolute values are different. Notably,
Ce is a neutron-capture element that forms during the asymptotic giant branch (AGB) evolutionary phase. The enhancement we see in Fig.~\ref{fig:Ce-V} could then be the result of the pollution by an AGB companion through super winds or Roche Lobe overflow.
Vanadium is enhanced by a slightly smaller amount in the YAR sample compared to the thick disc, which is confirmed by a similar result from the BAWLAS catalogue. 

\section{Extending to higher gravities}
\label{sec:general_YAR}
In this section, we explore the possible YAR candidates among the entire range of surface gravities. We selected candidates with $\rm \log g < 3.5$ (red clump and red giant stars) by applying the selection criterion described in Section \ref{subsec: YAR-defin} (age less than 4 Gyr and age uncertainty less than 3 Gyr). 
As mentioned in Section \ref{subsec:age}, the age dating method used in astroNN should not be used with dwarfs because the $\rm [C/N]$ abundance variation linked to the stellar mass occurs during the first dredge-up, after the main sequence phase.
For this reason, we directly identified dwarf straggler ($\rm 3.5 < log\, g < 4.45$) stars from their position relative to the main sequence turn-off (TO) point in the HRD (blue stragglers stars being located to the left and to the upper part of the main sequence TO). 
We additionally enlarged the original APOGEE DR17 sample from which the YAR are selected (see Section \ref{sec:data}), including stars selected for telluric lines (EXTRATARG==5). In order to obtain the most comprehensive sample possible, the selection of dwarf stragglers is subsequently made from this augmented sample.
\begin{figure}
\includegraphics[width=\hsize]{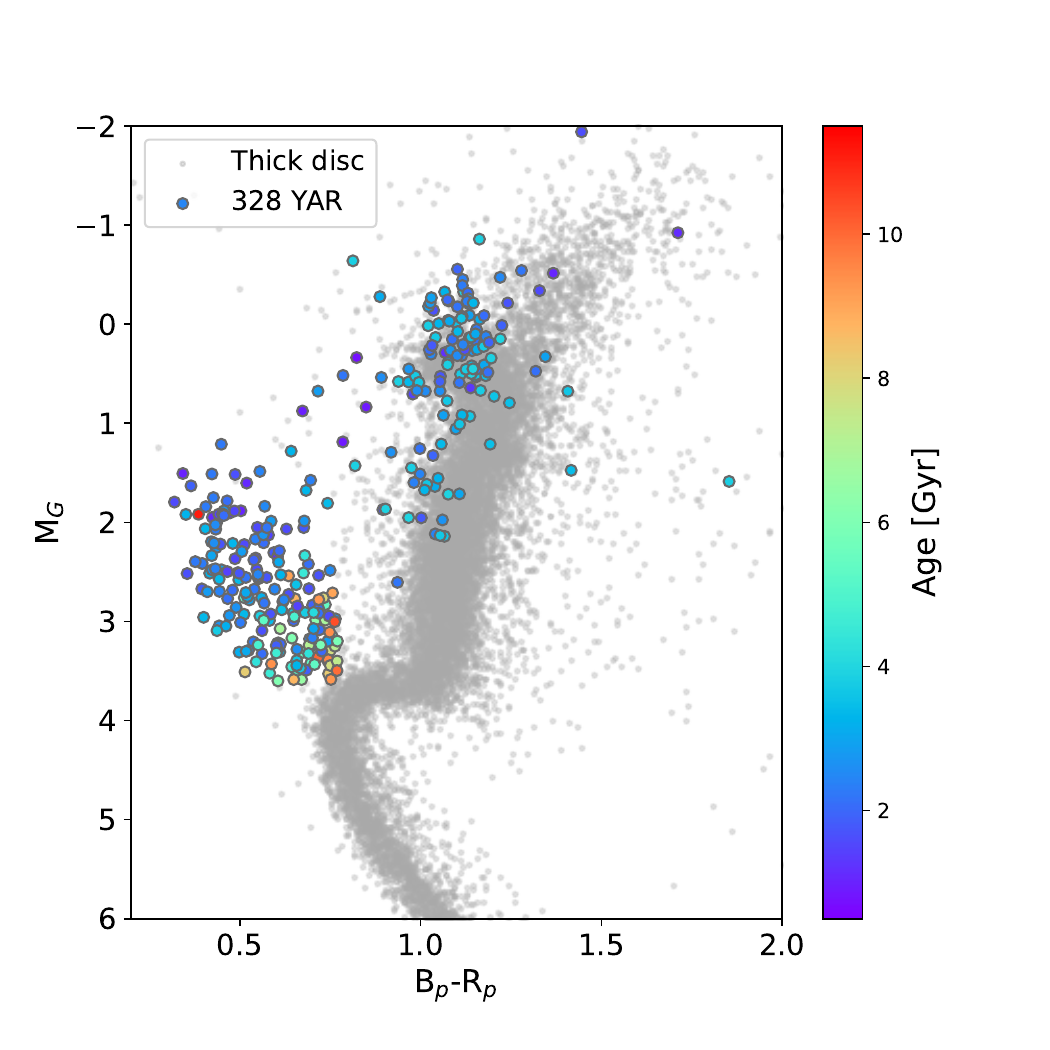}
\includegraphics[width=\hsize]{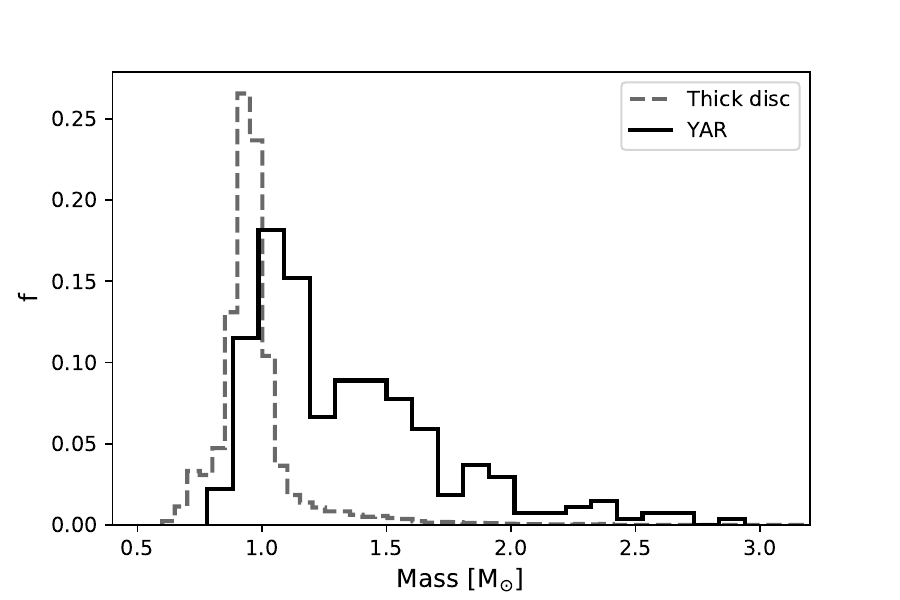}
\caption{Properties of 328 dwarf, red clump, and red giant straggler stars within 2 kpc from the Sun. \textit{Top}: Gaia HRD. YAR stars are colour-coded by their age.  Red giant stars and red clump YAR candidates with $\rm log\,g < 3.5$ are selected to have  an age less than 4 Gyr and an error on age determination less than 3 Gyr, while dwarfs stragglers candidates with $\rm 3.35 < log\,g < 4.45$ are selected from the position relative to the turn - off point in the HRD (see the text for details on the selection).}
Grey points are thick disc stars within 2 kpc from the Sun not selected on log g nor age. Photometry has been corrected for extinction and reddening using the reddening map from \citet{Lallement2019}. \textit{Bottom}: Mass histogram of YAR stars for which a mass determination is available from the StarHorse catalogue (270 objects, black histogram), compared with the thick disc sample of the upper panel (grey dashed histogram).
\label{dhr_stragglers_Gaia}
\end{figure}
Figure \ref{dhr_stragglers_Gaia} (top) shows the Gaia HRD of the sample of YAR stars selected within 2 kpc from the Sun, with points colour\text-coded by their age. The distance utilised to obtain the absolute magnitude are either the distance estimates provided in the catalogue for stars with $\rm \log g < 3.5$ or the inverse of the Gaia parallax for stars with $\rm \log g > 3.5$, astroNN distances overestimating the distances of the nearest stars. We choose to represents these nearby stars through the Gaia colour\text-magnitude diagram rather than the ($\rm M_{K0},\, T_{eff}$) HRD because good estimates of the extinction are available in visible bands within 2 kpc from the Sun, while the \texttt{AK$\text{\textunderscore}$TARG} in the K band given in the APOGEE DR17 is overestimated.
Because of the limit in distance, the number of giants is reduced, but the dwarf stragglers to the left and upper part of the thick disc TO (located at $\rm M_{G} = 3.6$ and $\rm (G_{BP} - G_{RP}) = 0.77$) are well visible. They represent around 57$\rm \%$ of the overall solar vicinity YAR sample and around 1$\rm \%$ of the solar vicinity thick disc sample. We note that our selection of dwarf stragglers includes stars that have a measured age of over 4 Gyr, as shown in Fig.~\ref{dhr_stragglers_Gaia} through the use of colour coding. The bottom plot of Fig.~\ref{dhr_stragglers_Gaia} shows the mass histogram of stars for which a mass is available in the StarHorse catalogue, illustrating the clear difference between the standard thick disc sample and the YAR sample. 
\begin{figure*}
\resizebox{\hsize}{!}{
\includegraphics[width=19cm]{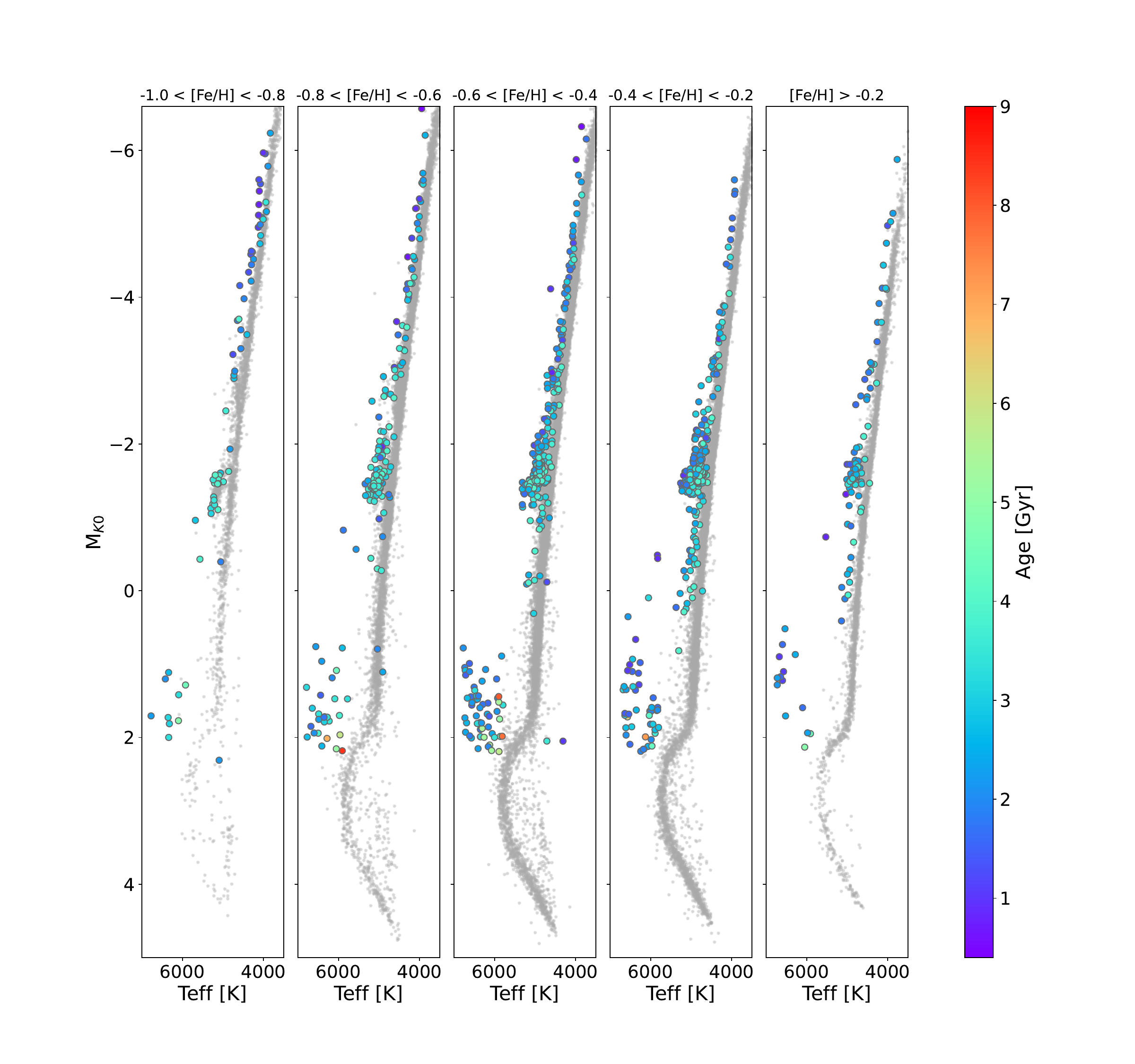}}
\caption{$\rm M_{K0}$ \text- Teff diagram of YAR candidate stars in the entire $\rm log\, g$ range in different metallicity intervals. Red giant stars and red clump YAR candidates  with $\rm \log\,g < 3.5$ are selected on the thick disc sequence to have an age $\rm <$ 4 Gyr and an error on age determination less than 3 Gyr. Dwarfs stragglers candidates with $\rm 3.5 < log\, g < 4.45$ are selected to be to the left and to the upper part of the TO ($\rm M_{K0} = 2.2$, $\rm Teff = 5800 $ [K]). The grey points are stars on the thick disc sequence and no selection on  $\rm log\, g$ nor age.}
\label{dhr_stragglers_all}
\end{figure*}
Figure~\ref{dhr_stragglers_all} shows the (T$_{\rm eff}$, M$_{\rm K0}$) HRD for the YAR candidates in the entire range of $\rm \log g$ and distances separated in different metallicity intervals.
As for the previous figure, we relied on astroNN distances, except for stars with $\rm 3.5 < \log g < 4.45$, for which we used the inverse of the Gaia parallaxes (given that astroNN distances overestimate the distances of the nearest stars). We also corrected the magnitudes for extinction using the AK$\textunderscore$TARG value given in the catalogue. 
We select the YAR stars, colour-coded by their age in Fig.~\ref{dhr_stragglers_all}, following the same method adopted to obtain the YAR solar vicinity sample of Fig. \ref{dhr_stragglers_Gaia}. In this case, the main sequence TO utilised to distinguish dwarfs stragglers is located at ($\rm M_{K0} = 2.2$, $\rm Teff = 5800 $ [K]).
The fraction of stragglers selected in each metallicity interval relative to the total number of stars up to 1 magnitude below the turn-off is  2.4$\rm \%$, 1.6$\rm \%$, 1.8$\rm \%$, 1.9$\rm \%,$ and 2.0$\rm \%$. Among these stars, the percentage of dwarfs ($\rm 3.5 < \log g < 4.45$) is 15.9$\rm \%$, 14.9$\rm \%$, 15.0$\rm \%$ 13.8$\rm \%$, and 13.1$\rm \%$. These numbers of course depend on the definition adopted for selecting YAR objects and should only be taken as indicative. This is lower than the number of BSSs found in the oldest open clusters, where this fraction is an increasing fraction of the cluster age, reaching values above 5$\rm \%$ for clusters older than 1 Gyr \citep{Rain2021}, but this is unsurprising given our conservative limit on the upper stellar age (4 Gyr) of our YAR stars.

\section{Discussion}\label{sec:discussion}
We show that YAR red giant stars identified as outliers to the $\rm age-[\alpha/Fe]$ relation have chemical (in term of $\rm [\alpha/Fe]$ and $\rm [Fe/H]$ distributions) and kinematical characteristics similar to those of the thick disc population over the same metallicity interval.
We show that these YAR are correctly identified as more massive stars in the HRD, being systematically hotter than standard thick disc stars and they overlap well with isochrone of 2 Gyr age. 
Based on this observational evidence, the YAR stars in our dataset are most probably stars originating from the thick disc that became stragglers being subjected to a mass increase by either mass transfer or collision and/or coalescence. Chemical abundances are our best tools for investigating distinct straggler formation scenarios, considering that scenarios leading to different chemical patterns depend on several factors.
The chemical characteristics of these objects show a possible small systematic departure ($\rm \sim$0.01 dex$\rm /$0.02 dex) from the standard thick disc population for some elements, such as O, Ca, Ti, and Al. The difference is larger for Na ($\rm \sim$ 0.1 dex) and is confirmed by Na abundances obtained by \citet{Hayes2022} (BAWLAS catalogue). We relate this to a possible conversion of Na in more massive stars. Figure \ref{fig:Na} highlights that the shift in Na between the YAR and thick disc populations (visible also in Fig.~\ref{fig:Al}) is produced from the YAR stars of mass $\rm > 1.2\, M_{\odot}$. Stars in this mass range can experience mixing, which can modify the Na abundance at the surface of the star \citep{Smiljanic2009, Smiljanic2012, Luck1994}.
Finally, Ce, a slow neutron-capture element formed during the AGB phase, shows also a small positive offset ($\rm \sim 0.1$ dex), compared to the standard thick disc population also found in the BAWLAS catalogue. Such enhancement is expected if these YAR stars are component of binary systems where the more massive companion has evolved to the AGB phase and transfer some of its mass to the lower mass companion.
Although these differences seem to be real, they are small and need to be confirmed. 

While the mass transfer scenario (by either stellar wind or Roche lobe overflow, RLO) is in agreement with these observational indices, it is possible that some of the YAR stars have been formed through collision or coalescence mechanisms -- for several reasons. First, little or no modification of stellar abundance is expected in the case of collision, as already mentioned \citep{Lombardi1995, Lombardi1996, Sills2001}, which may be the case for some of our stars. Second, the mass determinations shown on Figs.~\ref{fig:Mass_distrib} and \ref{dhr_stragglers_Gaia} suggest that the highest masses found may be difficult to achieve with mass transfer through stellar wind or RLO. 
Finally, recent studies of the dynamics of triple systems have shown that secular evolution of these systems may lead to collisions \citep{He_Petrovich_2018, Toonen2020, Toonen2022, Grishin2022} and could provide an efficient way to form stragglers in low-density environments such as the field. In this case, the evolution of the most massive component can destabilise the system and lead to the collision of the two closest components.
A merger or coalescence of close binaries can also occur and produce objects that are more massive than via the mass transfer or wind scenarios \citep{Sills2015}.  

\section{Conclusions}\label{conclusions}
In this work, we made use of the APOGEE DR17 survey and of the astroNN value-added catalogue to identify a sample of 
249 YAR red giants stars. We selected these objects to be $\alpha$\text-enhanced and younger than 4 Gyr.
In the first part of the paper, we characterise the YAR population, comparing the global properties of the YAR stars with a selected thick disc reference sample  (24,357 red giant stars). 
The relative position of the YAR stars in the (M$_{\rm K0}$, T$_{\rm eff}$) diagram indicates that they are more massive than the thick disc reference stars. This is also supported by the comparison of the mass distributions of the two samples.

Overall, the YAR stars present $\rm [\alpha/Fe]$ and $\rm [Fe/H]$ patterns which resemble those of the thick disc. However, we point out the presence of a small discrepancy in the $\rm \alpha$\text-distribution of about $\rm \sim 0.01$ dex: the YAR stars follow the thick disc trend, but are shifted to lower values of $\rm [\alpha/Fe],$ in particular at $\rm [Fe/H] > -0.4$. 

From the kinematical point of view, we demonstrate that the characteristics of our YAR sample stars are similar to those of the reference thick disc sample. Taking into account the described properties, we do record the YAR stars in our dataset as part of the thick disc population.
From this outlook these objects are most likely straggler stars (of the thick disc), as has  been previously suggested (references in Section \ref{sec:intro}).

We searched for candidates of accreted straggler stars in our dataset, picking out an halo-like population and excluding the so-called heated thick disc population. We obtained 25 candidates, however, their analysis led to inconclusive results. Presumably, this sample is not extensive enough to outline its global properties with confidence.

In the second part of the paper, we study the chemical patterns of  the YAR stars compared to the thick disc reference sample. 
The individual $\alpha$\text-elements investigated are O, Mg, Si, S, Ca, and Ti. In particular, Mg and O seem to be the main cause of the difference in the $\rm [\alpha/Fe]-[Fe/H]$ distribution between YAR and thick disc stars. Also, Ca and Ti exhibit a clear shift between the two samples, but opposite to the trend  for oxygen and magnesium. We find a clear offset in sodium of about $\rm \sim 0.1$ dex. The $\rm [Na/Fe]-mass$ distribution suggests that the enhanced Na abundance in the YAR stars  is linked to their increased masses. The enhancement could be related to some mixing phenomena, which could, in turn, depend on the mass of the stars. Lastly, we notice that the YAR stars tend to be more vanadium- and cerium-enhanced with respect to the standard thick disc population. The case of Ce is noteworthy, considering that s\text-neutron capture elements have been proposed as clues to the existence of mass transfers between an asymptotic giant branch star and a companion.

Finally, we report the results obtained after extending the YAR selection criterion to higher gravities in the APOGEE DR17 dataset. In particular, the dwarf YAR stars are located in the HRD in the same position where standard BSSs are observed.  The fraction of dwarf stragglers of our dataset is lower than the number of BSSs found in the oldest (age $\rm >$ 1 Gyr) open clusters, but this could be due to our conservative cut on age. 
In light of these investigations, the formation pathway that is most consistent with our results is that of  mass acquisition via mass transfer in a binary system. However, it is possible that the most massive straggler stars of our sample have been produced by collision or coalescence. 

\begin{acknowledgements}
The authors would like to thank Ted Mackereth and Henry Leung for helpful comments on the age determination from the $\rm [C/N]$ ratio, and Christian Hayes for providing the BAWLAS catalogue. 
The authors acknowledge the support of the French Agence Nationale de la Recherche (ANR), under grant ANR-13-BS01-0005 (project ANR-20-CE31-0004-01 MWDisc).
This work has made use of data from the European Space Agency (ESA) mission Gaia (https://www.cosmos.esa.int/gaia), processed by the Gaia Data Processing and Analysis Consortium (DPAC, https://www.cosmos.esa.int/web/gaia/dpac/consortium). Funding for the DPAC has been provided by national institutions, in particular the institutions participating in the Gaia Multilateral Agreement.
This research made use of Astropy, a community-developed core Python package for Astronomy (Astropy Collaboration, 2018).
Funding for the Sloan Digital Sky 
Survey IV has been provided by the 
Alfred P. Sloan Foundation, the U.S. 
Department of Energy Office of 
Science, and the Participating 
Institutions. 

SDSS-IV acknowledges support and 
resources from the Center for High 
Performance Computing  at the 
University of Utah. The SDSS 
website is www.sdss.org.

SDSS-IV is managed by the 
Astrophysical Research Consortium 
for the Participating Institutions 
of the SDSS Collaboration including 
the Brazilian Participation Group, 
the Carnegie Institution for Science, 
Carnegie Mellon University, Center for 
Astrophysics | Harvard \& 
Smithsonian, the Chilean Participation 
Group, the French Participation Group, 
Instituto de Astrof\'isica de 
Canarias, The Johns Hopkins 
University, Kavli Institute for the 
Physics and Mathematics of the 
Universe (IPMU) / University of 
Tokyo, the Korean Participation Group, 
Lawrence Berkeley National Laboratory, 
Leibniz Institut f\"ur Astrophysik 
Potsdam (AIP),  Max-Planck-Institut 
f\"ur Astronomie (MPIA Heidelberg), 
Max-Planck-Institut f\"ur 
Astrophysik (MPA Garching), 
Max-Planck-Institut f\"ur 
Extraterrestrische Physik (MPE), 
National Astronomical Observatories of 
China, New Mexico State University, 
New York University, University of 
Notre Dame, Observat\'ario 
Nacional / MCTI, The Ohio State 
University, Pennsylvania State 
University, Shanghai 
Astronomical Observatory, United 
Kingdom Participation Group, 
Universidad Nacional Aut\'onoma 
de M\'exico, University of Arizona, 
University of Colorado Boulder, 
University of Oxford, University of 
Portsmouth, University of Utah, 
University of Virginia, University 
of Washington, University of 
Wisconsin, Vanderbilt University, 
and Yale University.
\end{acknowledgements}

\bibliographystyle{aa} 
\bibpunct{(}{)}{;}{a}{}{,}             
\bibliography{overleaf/Bibliography} 

\end{document}